\documentclass[aps,prb,twocolumn,showpacs]{revtex4-1}

\usepackage{graphicx}
\DeclareGraphicsExtensions{.png}
\DeclareGraphicsExtensions{.jpeg}

\usepackage{xcolor}
\usepackage{amsmath}
\usepackage{amssymb}
\usepackage{bbm}

\usepackage{dcolumn}% Align table columns on decimal point
\usepackage{bm}% bold math
\usepackage{hyperref}
\usepackage[applemac]{inputenc}
\usepackage[american]{babel}
\usepackage{latexsym}
\usepackage{float}

\begin{document}

\title{Modulation of Kekulé adatom ordering due to strain in graphene}

\index{J. González-Árraga}
\index{F. Guinea}
\index{P. San-Jose}

\author{L. González-Árraga$^{1}$, F. Guinea$^{1,2}$ and P. San-Jose$^{3}$}
\affiliation{${}^1$IMDEA Nanociencia, Calle de Faraday 9, 28049 Madrid, Spain}
\affiliation{${}^2$Department of Physics and Astronomy, University of Manchester, Manchester M13 9PL, United Kingdom}
\affiliation{${}^3$Materials Science Factory, ICMM-CSIC, Sor Juana Ines de La Cruz 3, 28049 Madrid, Spain}

\date{\today}

%%%%%%%%%%%%%%%%%%%%%%%%%%%%%%%%%%%%%%%%%%%%%%%%%%%%%%%%%%%%%%%%%%%%%%%%%%%%%%%%
\begin{abstract}
Intervalley scattering of carriers in graphene at `top' adatoms may give rise to a hidden Kekulé ordering pattern in the adatom positions. This ordering is the result of a rapid modulation in the electron-mediated interaction between adatoms at the wavevector $\bm K-\bm K'$, which has been shown experimentally and theoretically to dominate their spatial distribution. Here we show that the adatom interaction is extremely sensitive to strain in the supporting graphene, which leads to a characteristic spatial modulation of the Kekulé order as a function of adatom distance. Our results suggest that the spatial distributions of adatoms could provide a way to measure the type and magnitude of strain in graphene and the associated pseudogauge field with high accuracy. 
%Conversely, atomic-scale control of the relative position of top adatoms may be achieved through strain engineering.
\end{abstract}

\maketitle

%%%%%%%%%%%%%%%%%%%%%%%%%%%%%%%%%%%%%%%%%%%%%%%%%%%%%%%%%%%%%%%%%%%%%%%%%%%%%%%%

Much of the rich physics of graphene  stems from the peculiarities of its intrinsic electronic structure, such as its gapless Dirac spectrum, the chirality of its carriers, or the emergence of pseudogauge fields as a result of inhomogeneous strains~\cite{Neto:RMP09,Das-Sarma:RMP11, Amorim:PR16}. These are all `intra-valley' properties, defined independently within valleys $\bm K$ and $\bm K'$. They are responsible for e.g. graphene's high mobilities~\cite{Bolotin:SSC08}, Klein tunneling~\cite{Beenakker:RMP08}, the valley-Hall effect~\cite{Gorbachev:S14} or the emergence of topologically protected boundary states in bilayers~\cite{Martin:PRL08,San-Jose:PRB13}. They remain robust as long as valley symmetry is preserved, i.e. as long as any perturbation or disorder present in the sample acts symmetrically on the two sublattices of the crystal.
% and do not couple $\bm K$ and $\bm K'$. 
%Such is the case of charge puddles created by typical substrates, or spatially smooth potentials from electrostatic gates. 
Atomic-like defects are one important type of perturbation that does not in general preserve valley symmetry, and allows for scattering events with an intervalley $\Delta \bm K=\bm K-\bm K'$ momentum transfer ($\hbar=1$)~\cite{Chen:PRL09}. 

Intervalley scattering may be important at the edges of a generic graphene flake~\cite{Cresti:PRB09,Libisch:NJP12}, at substitutional dopants~\cite{Lawlor:PRB13,Zhao:S11,Pasti:17}, or at certain adatoms~\cite{Pachoud:PRB14} that adsorb to graphene in a `top' configuration (i.e. adsorbed atop individual carbon atoms), such as Fluor~\cite{Nair:S10} or Hydrogen~\cite{Gonzalez-Herrero:S16}, thereby breaking sublattice symmetry. Despite destroying the chiral nature of carriers in graphene, intervalley scattering is also fundamentally interesting in its own right~\cite{Morpurgo:PRL06}, and can actually become a powerful tool, particularly for graphene functionalization. It is crucial for the engineering of enhanced spin-orbit couplings~\cite{Pachoud:PRB14} and finite bandgaps in graphene via decoration with adatoms~\cite{Cheianov:PRB09, Balog:NM10, Cheianov:EEL10,Wang:SR15}, by the effect of a crystalline substrate~\cite{Wallbank:PRB13a,Jung:NC15,Zhou:NM07}, or through electron-phonon interaction~\cite{Iadecola:PRL13}. 

Here we focus on another striking effect of intervalley scattering, the unique ordering mechanism of top adatoms~\cite{Balog:NM10} and similar atomic like defects~\cite{Lv:SR12, Gutierrez:NP16} in graphene. Ordering results from the electron-mediated interactions between defects as graphene quasiparticles scatter between them~\cite{Shytov:PRL09,Cheianov:SSC09,LeBohec:PRB14}. Scattering at adatoms locally modifies the electronic density of states in graphene, which gives rise to Friedel oscillations \cite{Cheianov:PRL06,Lawlor:PRB13} and to a change in the total electronic energy that depends on the distance between adatoms.
This gives rise to a fermionic analogue of the Casimir force~\cite{Zhabinskaya:PRA08}, and has been shown to be the dominant contribution in the interaction between graphene adatoms~\cite{Solenov:PRL13}. It leads to the self-organization of atomic defects and adatoms at different levels, including sublattice ordering~\cite{Cheianov:EEL10,LeBohec:PRB14}, Kekulé ordering~\cite{Cheianov:SSC09}, and spatial clustering~\cite{Shytov:PRL09}. Kekulé ordering, recently demonstrated in experiment~\cite{Gutierrez:NP16}, is probably the most striking of these. In this work we show that electron-mediated Kekulé ordering is extremely sensitive to elastic strains in the underlying graphene. The connection arises from the effect of strain-induced pseudogauge fields on intervalley scattering, and could provide a sensitive way to measure strains through adatom distributions, or conversely to control Kekulé ordering of adatoms through strain engineering.

%%%%%%%%%%%%%%%%%%%%%%%%%%%%%%%%%%%%%%%%%%%%%%%%%%%%%%%%%%%%%%%%%%%%%%%%%%%%%%%%%%%%
\begin{figure}
   \centering
   \includegraphics[width=\columnwidth]{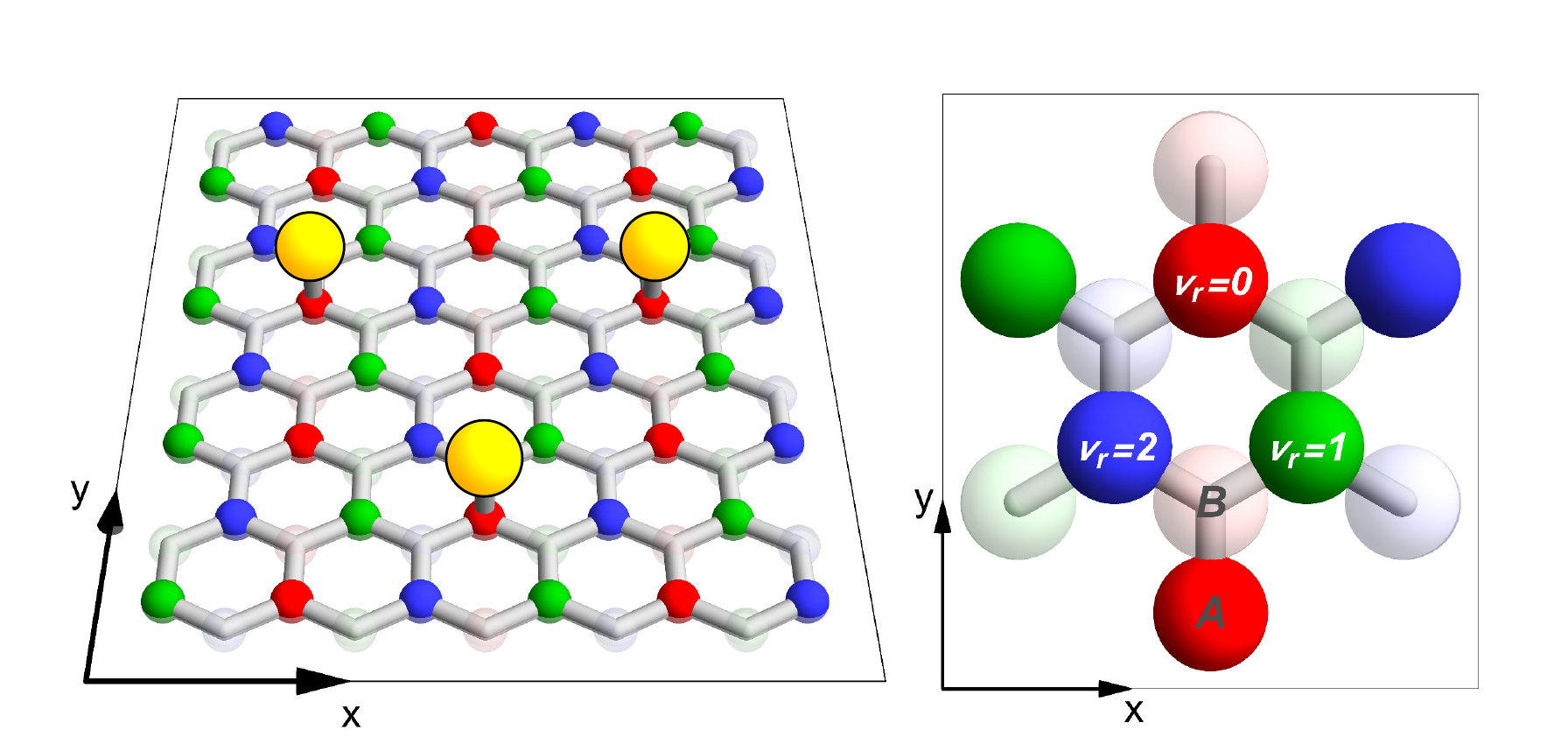}
   \caption{Sketch of the Kekulé index $\nu_{\bm r}=0,1,2$ (red, green, blue) of sites on each (A/B) sublattice in graphene. Hidden Kekulé ordering of top-adatoms (in yellow) corresponds to adsorption on sites with equal $\nu_{\bm r}$ (here $\nu_{\bm r}=0$), as a result of adatom interaction mediated by carriers in graphene that undergo intervalley scattering. Sublattice correlation of adatoms may also arise from the same interaction.}
   \label{fig:sketch}
\end{figure}
%%%%%%%%%%%%%%%%%%%%%%%%%%%%%%%%%%%%%%%%%%%%%%%%%%%%%%%%%%%%%%%%%%%%%%%%%%%%%%%%%%%%

%Intervalley momentum transfer $\Delta \bm K$ upon scattering on an adatom introduces a rapid variation of the electron-mediated adatom interaction typically of the form $\sim -\cos^2\left(\Delta \bm K\cdot\bm r\right)$. Kekulé ordering corresponds to adatoms arranging at distances $\bm r$ such that this modulation is minimized, i.e. $\Delta \bm K\cdot r=2\pi n$ for some integer $n$.

%The adatom interaction potential contains rapid variations on the atomic scale, precisely due to intervalley scattering of quasiparticles at the adatoms. Indeed, intervalley momentum transfer endows the interaction potential (and also the RKKY exchange interaction between top magnetic impurities\cite{Sherafati:PRB11}) with a rapid $\sim \cos^2\left(\Delta \bm K\cdot\bm r\right)$ modulation, known as a Kekulé pattern, which exhibits a minimum every three graphene unit cells, see Fig. \ref{fig:Kekule}. The Kekulé modulation of the interaction produces a long range `hidden Kekulé order' in an otherwise apparently random distribution of top adatoms on graphene. 

Consider a top adatom on sublattice $\sigma=$A,B of a graphene unit cell centered at $\bm r=n_1 \bm a_1+n_2 \bm a_2$ ($\bm a_i$ are graphene's lattice vectors with $|\bm a_i|=a_0$ and $|\Delta \bm K|=8\pi/\sqrt{3}a_0$). One may classify such adatom by the sublattice $\sigma$ and an integer Kekulé index $\nu_{\bm{r}}$, such that $ \Delta \bm K\cdot \bm r =2\pi\nu_{\bm{r}}/3 + 2\pi n$ for some integer $n$, i.e.
\begin{equation}
\nu_{\bm{r}} = \frac{\Delta \bm K\cdot \bm r}{2\pi/3} \!\!\!\mod 3= (n_1-n_2)\!\!\!\mod 3 = 0,1,2.
\end{equation}
These three possibilities are color-coded as `red', `green' and `blue' here, and are shown in Fig. \ref{fig:sketch} for one of the graphene sublattices.
Hidden Kekulé order \cite{Gutierrez:NP16} consists of collections of top adatoms or atomic defects which minimise their quasiparticle-mediated interaction energy by adopting the same values of $\nu_{\bm r}$, and (possibly) the same value of $\sigma$, see yellow adatoms in Fig. \ref{fig:sketch}. We now describe the mechanism that gives rise to Kekulé ordering, and then analyse how it is affected by the presence of elastic strains.

%Very clear evidence of this kind of ordering has been recently obtained experimentally for a certain type of ``vacancy adsorbate", and has been shown to be remarkably robust up to room temperature.

The interaction between two adatoms on graphene has various contributions, including local elastic deformations of graphene around adatoms, direct overlap of adatom orbitals, direct Coulomb interactions (monopolar or multipolar) and interactions mediated by scattering of quasiparticles in graphene. Of these, only the last two are relevant in realistic conditions~\cite{Solenov:PRL13}, with the latter dominating the interaction of neutral adatoms. Direct Coulomb interactions are rather simple, and do not produce any Kekulé ordering, so we will concentrate on the far richer properties of the electron-mediated interaction potential $U(\bm r)$. We model the graphene-adatom system in a tight-binding approximation,
\begin{equation}\label{H}
H=-t\sum_{\langle i,j\rangle} c^\dagger_ic_j + \epsilon_0\sum_k d^\dagger_kd_k - t'\sum_k (d^\dagger_kc_k + c^\dagger_kd_k), 
\end{equation}
where $c_i$ are graphene $\pi$ orbitals, and $d_k$ are adatom states, located at positions $\bm{r}_k$, and coupled to a single $c_k$ state in graphene (top configuration). Consider for simplicity only two adatoms in the system $k=1,2$ on sublattices $\sigma_1$ and $\sigma_2$ at a distance $\bm{r}=\bm{r}_2-\bm{r}_1$. The interaction potential $U_{\sigma_1\sigma_2}(\bm r)$ can be written \cite{Hyldgaard:JOPCM00,Shytov:PRL09,LeBohec:PRB14} as the total energy of all the electrons in the system, as they adjust to the presence of the adatoms,
\begin{eqnarray}
U_{\sigma_1\sigma_2}(\bm{r})&=&\int d\omega\,  (\rho(\omega)-\rho_\infty(\omega))\,\omega f(\omega)\\
\rho(\omega)&=&-\frac{1}{\pi}\mathrm{Tr}\left[G^{\sigma_1}_1(\omega)+G^{\sigma_2}_2(\omega)+\int d^2 r' G(\bm r', \bm r'; \omega)\right].\nonumber
\end{eqnarray}
Here $G$ and $G^{\sigma_k}_k$ are the full, retarded Green function of graphene and the two adatoms, respectively, and $f(\omega)$ is the Fermi function (for concreteness, zero temperature and zero filling are assumed from now on). The potential depends implicitly on the adatom distance $\bm{r}$, and contains fast spatial harmonics $\sim \cos(\Delta \bm K\cdot\bm{r})$ due to the interference of $\bm K$ and $\bm K'$ that results from intervalley scattering. The reference density of states $\rho_\infty(\omega)$ is defined as the limit for $\bm{r}\to\infty$, so that $U(\bm{r}\to\infty) = 0$.

%%%%%%%%%%%%%%%%%%%%%%%%%%%%%%%%%%%%%%%%%%%%%%%%%%%%%%%%%%%%%%%%%%%%%%%%%%%%%%%%%%%%
\begin{figure}
   \centering
   \includegraphics[width=\columnwidth]{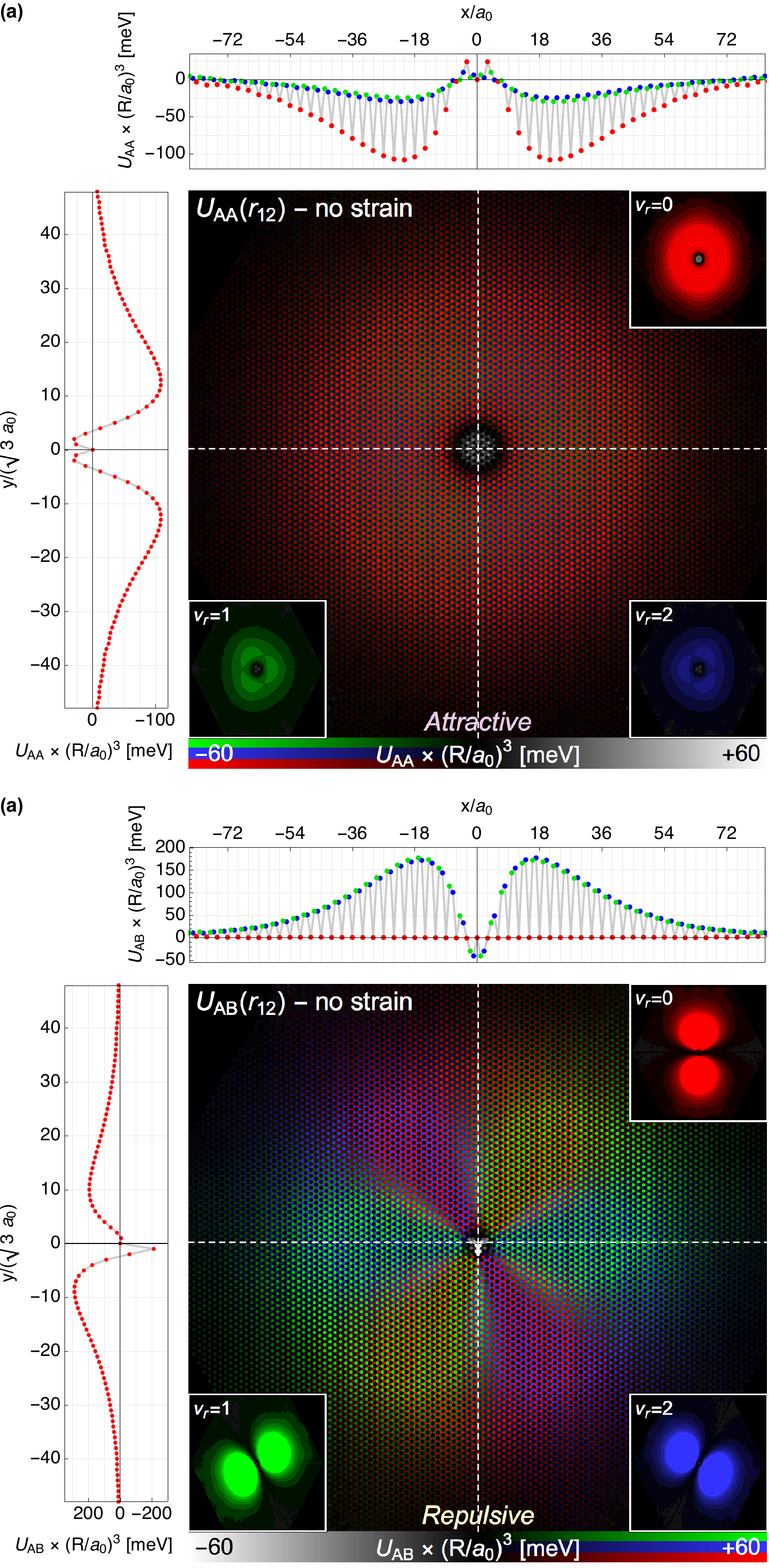}
   \caption{Interaction potential between two adatoms (with $\epsilon_0=-0.15 t$) on unstrained graphene, weakly attached ($t'=0.7 t$) to the same (a) and to different (b) subblattices. The interaction potential results from the scattering of graphene carriers at the adatom sites. The potential is attractive for equal sublattice, and repulsive otherwise. By color-coding the potential according to the Kekulé character $\nu_{\bm r}$ of the separation vector $\bm r=\bm r_2-\bm r_1$, we see that at equilibrium the two adatoms rest on the same sublattice, at sites with equal Kekulé character ($\nu_{\bm r}=0 \mathrm{(red)} \Rightarrow \nu_{\bm r_{1}}=\nu_{\bm r_{2}}$). Regardless of the radial dependence of the interaction potential [multiplied here by $(|\bm r|/a_0)^3$ for visibility], its Kekulé components (colored insets) satisfy the simple forms in Eqs. (\ref{UAA},~\ref{UAB}), up to small corrections from higher angular harmonics.}
   \label{fig:nostrain}
\end{figure}
%%%%%%%%%%%%%%%%%%%%%%%%%%%%%%%%%%%%%%%%%%%%%%%%%%%%%%%%%%%%%%%%%%%%%%%%%%%%%%%%%%%%

%%%%%%%%%%%%%%%%%%%%%%%%%%%%%%%%%%%%%%%%%%%%%%%%%%%%%%%%%%%%%%%%%%%%%%%%%%%%%%%%%%%%
\begin{figure*}
   \centering
   \includegraphics[width=\textwidth]{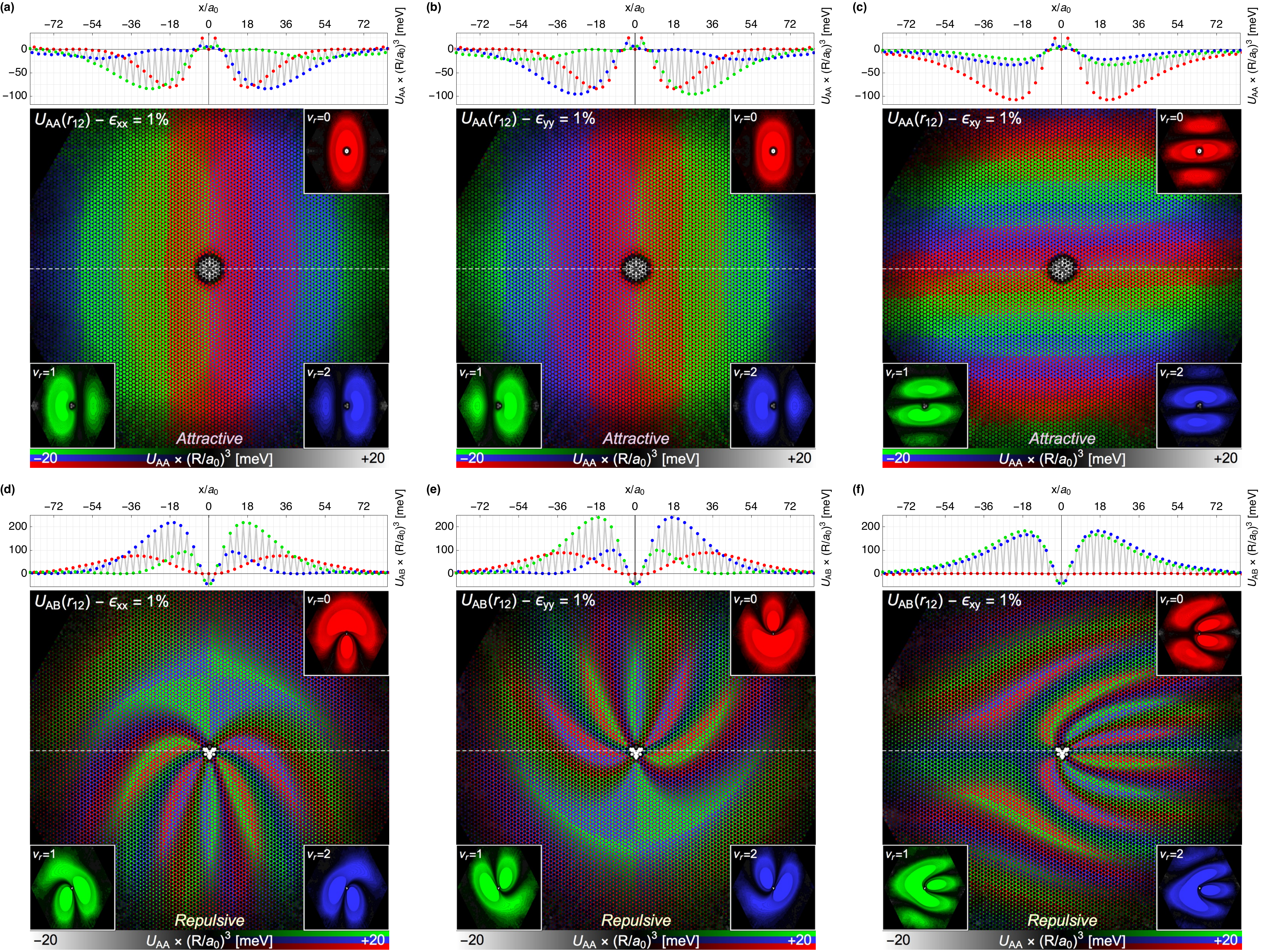}
   \caption{Interaction potential between two adatoms as in Fig. \ref{fig:nostrain}, with a 1\% uniform strain present in graphene, either uniaxial $\epsilon_{xx}$(a,d), uniaxial $\epsilon_{yy}$ (b,e), or a uniform shear $\epsilon_{xy}$ (c,f). Equal sublattice configurations is still preferred, but the Kekulé character acquires a modulation with distance that reflects the change of the intervalley separation $\bm K-\bm K'$ by a pseudogauge field $2\bm A\sim(\epsilon_{xx}-\epsilon_{yy}, -2\epsilon_{xy})$.}
   \label{fig:strain}
\end{figure*}
%%%%%%%%%%%%%%%%%%%%%%%%%%%%%%%%%%%%%%%%%%%%%%%%%%%%%%%%%%%%%%%%%%%%%%%%%%%%%%%%%%%%

We computed $U(\bm{r})$ numerically to all orders in the coupling $t'$, as described in the Appendix~\ref{ap:method}. In the weak coupling limit the results agree with analytical expressions for $U$ that have been obtained in the literature for unstrained graphene \cite{LeBohec:PRB14}. It was shown, using a simplified adatom model, that in the limit of weakly coupled adatoms, $U$ exhibit a Kekulé modulation given by
\begin{eqnarray}
U_{AA}(\bm r)&=&U_{BB}(\bm r)\approx v_{AA}\left(|\bm r|\right) \cos^2\left(2\pi\nu_{\bm r}/3\right), \label{UAA}\\
U_{AB}(\bm r)&=&U_{BA}(\bm r)\approx v_{AB}\left(|\bm r|\right) \sin^2\left(2\pi\nu_{\bm r}/3+\phi_{\bm r}\right).\label{UAB}
\end{eqnarray}
Here $\phi_{\bm r}$ is the angle between $\Delta\bm K$ and $\bm r=\bm r_2-\bm r_1$. $v_{AA}(r)$ and $v_{AB}(r)$ are smooth functions of inter-adatom distance, that in absence of dissipation fall as $1/r^3$ at long distances. The sign of $v_{\sigma\sigma'}(r)$ is controlled by the adatom coupling $t'$, and the interaction strength and decay with $r$ is strongly affected by inelastic processes in graphene (see Appendices \ref{ap:dissipation} and \ref{ap:strong} for further discussion). This results in a rich and partially tuneable interaction phenomenology between adatoms. Perhaps the most relevant property, however, is that for weak couplings, $v_{AA}(r)$ is attractive and $v_{AB}(r)$ is repulsive, while for strong coupling the opposite is true. Hence, adatoms will exhibit ferro- or antiferro-like order in the sublattice quantum number, depending on how strongly coupled they are to graphene. The Kekulé factor, in contrast, is much more universal, and the expressions in Eqs. (\ref{UAA},\ref{UAB}) remain qualitatively correct regardless of coupling strength. Corrections come in the form of weaker harmonics in the angular coordinate $\phi_{\bm r}$. The Kekulé factor effectively produces six different, spatially-smooth potential components $U_{\sigma_1,\sigma_2, \nu_{\bm r_1},\nu_{\bm r_2}}$ for each combination of $\sigma,\sigma'$ and $\nu_{\bm r_1},\nu_{\bm r_2}$. In the following, we will focus on these components to reveal the favored Kekulé and sublattice ordering in each case.

Numerical results for $U_{AA}(\bm r)$ and $U_{AB}(\bm r)$ without strain are shown in Fig. \ref{fig:nostrain}, panel (a) and (b) respectively. Parameters are chosen in the weak coupling regime, which corresponds to the phenomenology seen in the experiment of Ref. \onlinecite{Gutierrez:NP16}. In the insets we show the three Kekulé components corresponding to the three values of $\nu_{\bm r}=\nu_{\bm r_1-\bm r_2}=0,1,2$ (red, green and blue). The main panels show all the Kekulé components together in real space, but are plotted
so as to emphasize the color of the most (least) favored Kekulé component at each adatom distance for $U_{AA}$ ($U_{AB}$). 
%
%Distances $\bm r$ in main panel of (a) are plotted using a red, green or blue color scale according to their Kekulé character $\nu_{\bm r}=0,1,2$. Also, 
Points with a more negative potential $U_{AA}(\bm r)$ are rendered last in panel (a), so that the most visible color of a given point corresponds to the Kekulé character of the potential minimum. For $U_{AB}(\bm r)$ in panel (b) we use the opposite rendering order, so that the Kekulé character of points with the strongest repulsion is the most visible. Cuts of the potential along the vertical and horizontal directions are also included.
We note that $U_{AA}$ at $\nu_{\bm r}=0$ (red) is the most attractive potential component [top-right inset in panel (a)].  In the chosen parameter regime the electron-mediated potential favors isotropic configuration of adatoms on the same sublattice $\sigma$ and with equal Kekulé index. Note that the angular profile of all potential components follows Eqs. (\ref{UAA},\ref{UAB}).

%In a sample with many adatoms one therefore expects an isotropic distribution of adatoms on the same sublattice $\sigma_1=\sigma_2$, and with the same Kekulé index $\nu_{\bm r_1}=\nu_{\bm r_2}$.

%We study here the Kekulé ordering of top adatoms in the presence of homogeneous strain in graphene. The effect of a strain tensor $\epsilon_{ij}$ on graphene carriers is known to be equivalent to an opposite shift of the $\bm K$ and $\bm K'$ points by a pseudogauge field $\bm A$ that is linear if $\epsilon_{ij}$. In the case of homogeneous strain, this is of no consequence to intra-valley physics, as it can be gauged away. It has, however, a strong impact in intervalley scattering, since the Kekulé momentum transfer changes to $\Delta\bm K+2\bm A$. We show that, as a consequence, homogeneous strain induces a precession in space in the hidden Kekulé ordering of adatoms, and could thus be used to control adatom position and ordering on the atomic scale. Conversely, by measuring the Kekulé alignment of adatom pairs as a function of their relative distance it is possible to infer the magnitude of the strain of the underlying graphene sample to very high accuracy, and to easily distinguish uniaxial from shear strain types.

We now consider the same problem in the presence of uniform strain $\epsilon_{ij}$ in graphene, such that the position of each carbon atom $\bm r_i$ becomes $\bm r_i+\bm \epsilon\cdot\bm r_i$. The distortion is incorporated into the tight-binding description of Eq. \eqref{H} by making the hopping $t$ depend on the carbon-carbon distance as $t_{\bm \epsilon}=t\exp\left[-\beta(|\bm r_i-\bm r_j|/a_0-1)\right]$, where $\beta\approx 3$. For realistic strains, this shifts the $\bm K$ and $\bm K'$ valleys by an opposite pseudogauge vector 
\begin{equation}
\bm A = \pm \frac{2\beta}{\sqrt{3}a_0}(\epsilon_{xx}-\epsilon_{yy},-2\epsilon_{xy})
\label{A}
\end{equation}
(the $\hat{x}$ axis corresponds here to the zigzag direction). In the case of homogeneous strain, this pseudogauge potential is of no consequence to intra-valley physics, as it can be gauged away. It has, however, a strong impact in intervalley scattering, since the Kekulé momentum transfer changes to $\Delta\bm K+2\bm A$. Consequently, it would be natural to expect intervalley-dependent quantities such as $U_{\sigma\sigma'}(r)$ to exhibit signatures of a uniform strain. The weak-coupling Kekulé should then become
\begin{eqnarray}
U_{AA}(\bm r)&\approx& v_{AA}\left(|\bm r|\right) \cos^2\left(2\pi\nu_{\bm r}/3+2{\bm A}\cdot{\bm r}\right), \label{UAAs}\\
U_{AB}(\bm r)&\approx& v_{AB}\left(|\bm r|\right) \sin^2\left(2\pi\nu_{\bm r}/3+2{\bm A}\cdot{\bm r}+\phi_{\bm r}\right).\label{UABs}
\end{eqnarray}
This expectation is indeed confirmed by our numerical simulations. Figure \ref{fig:strain} shows the modified potential $U_{AA}$ (panels a-c)  and $U_{AB}$ (panels d-f) for the same parameters of Fig. \ref{fig:nostrain} under an uniform 1\% uniaxial strain along $x$ and $y$ directions, and a 1\% uniform shear strain. We concentrate on the $U_{AA}(\bm r)$ potential, as the $U_{AB}$ remains repulsive and is thus irrelevant for the equilibrium adatom configurations (see Appendix \ref{ap:strong} for additional results in the case of strong coupling). The equal-sublattice configuration is still the most stable one in the presence of strain in this regime. One immediately observes, however, a new spatial modulation in each of the Kekulé components that is linear in $\epsilon_{ij}$. While a uniform Kekulé adatom configuration $\nu_{\bm r}=0$ was favored in the case without strains, a 1\% strain makes the potential minimum change Kekulé character with distance, precessing between $\nu_{\bm r}=0,1,2$ (red, green, blue) as the two adatoms are separated (see vertical/horizontal stripes in Figs \ref{fig:strain}(a-c)). This type of precessing interaction is reminiscent of the Dzyaloshinskii-Moriya exchange interactions in chiral magnets~\cite{Dzyaloshinsky:JPCS58,Moriya:PR60}, responsible for the formation of skyrmion spin structures \cite{Nagaosa:NN13}, although here it operates in the Kekulé instead of the spin sector.

The spatial modulation is consistent with the form of $\bm A$ given in Eq. \eqref{A}. Uniaxial strain $\epsilon_{xx}$ and $\epsilon_{yy}$ along the $x$ and $y$ directions both modulate the Kekulé character along the $x$ direction, albeit in an opposite sequence order. In contrast, a shear strain $\epsilon_{xy}$ creates a modulation along the $y$ direction, with a period that is half that of the uniaxial strain. The modulation period is given by $\pi/|6\bm A|$, i.e. around 3-4 nm for 1\% of uniaxial strain. 

For a large ensemble of adatoms, the Kekulé orientation of domains should also exhibit a spatial modulation. A given adatom will align its Kekulé index to nearby adatoms, with which interaction is strongest. However, the long-range coherence of Kekulé domains will be controlled by the long-range component of the interaction, so striped Kekulé domains are expected to arise even under weak uniform strains. This requires sufficiently long-range interactions such as those observed in the experiment of Gutierrez \emph{et al.}~\cite{Gutierrez:NP16} (Kekulé domain sizes in the tens of nanometers and above, substantially greater than modulation periods at 1\% strains). In such cases the spatial modulation of Kekulé alignement is expected to show a high sensitivity to the magnitude and type (uniaxial/shear) of strains in the sample. 

We have concentrated here on the simplest case of a point-like adatom in a top configuration. More complex adsorbates, such as larger molecules or adatoms in different stacking configurations (hollow and bridge) should be expected to result in different interaction potentials. Likewise, the inclusion of further physical ingredients, such as electronic interactions and adatom magnetism could extend the results presented here. We have explored a number of these extensions in Appendix \ref{ap:interactions} (strong coupling, onsite interactions, adatom magnetisation and RKKY exchange \cite{Cheianov:PRB09, Black-Schaffer:PRB10, Sherafati:PRB11,Kogan:PRB11}). While quantitative differences where found, they were mostly confined to the range and sign of the different smooth Kekulé components $v_{\sigma\sigma'}(r)$. The Kekulé modulation of the potential and its dependence with strain, Eqs. (\ref{UAAs},~\ref{UABs}), remain mostly unchanged. The fundamental connection between Kekulé order and strain is thus found to be universal, and is one of the most striking manifestations of uniform strains in graphene. 
%Non-uniform strains are also expected to affect adatom interactions, although in such case a competition between the pseudomagnetic length and adatom distance becomes possible, and its implications in equilibrium adatom configurations remains to be explored. 

\acknowledgements

L. G-A. and F. G. acknowledge the financial support by Marie-Curie-ITN Grant No. 607904-SPINOGRAPH. P.S-J. acknowledges financial support from the Spanish Ministry of Economy and Competitiveness through Grant No. FIS2015-65706-P (MINECO/FEDER).

\appendix

\section{Interaction potential between two adatoms}
\label{ap:method}

The total interaction energy between two identical adatoms in a top configuration on a graphene monolayer can be decomposed in several contributions. Solenov \emph{et al.} showed~\cite{Solenov:PRL13} that the dominant contributions are reduced to two: the electrostatic repulsion and the interaction mediated by electron scattering in graphene. The first may be present even for charge-neutral adatoms in multipolar form, but is otherwise rather simple. The second results in a much richer structure to the interaction, and has been shown to strongly dominate the ordering of impurities in some situations. 

The main feature of the electron-mediated interaction $U(\bm{r})$ between adatoms on graphene is that, by virtue of the strong intervalley ($\bm{K}\leftrightarrow\bm{K}'$) scattering at top-adatoms, it is rapidly modulated on the atomic lattice as $U(\bm{r}_i)\sim -\cos[(\bm{K}-\bm{K}')\cdot \bm{r}_i]$, which yields a characteristic Kekulé pattern in the potential minima. This was recently shown to produce a robust "hidden" Kekulé ordering of certain types of impurities that survives even at room temperature \cite{Gutierrez:NP16}.

Here we develop a derivation of the potential $U(\bm{r})$ using a simplified model for the adatoms. We describe graphene quasiparticles using a nearest-neighbour tight-binding model on the honeycomb lattice. The corresponding spin-degenerate Bloch Hamiltonian reads
\[
H_0(\bm k)=-t\left(
\begin{array}{cc}
0 & 1+e^{-i\bm{k}\cdot \bm{a}_1}+e^{-i\bm{k}\cdot \bm{a}_2}\\
1+e^{i\bm{k}\cdot \bm{a}_1}+e^{i\bm{k}\cdot \bm{a}_2} & 0
\end{array}
\right)
\label{Dirac}
\]
where $\bm{a}_i$ are the lattice vectors, and the matrix is expressed in  sublattice, space, which we denote by $\alpha=A,B$.  The Hamiltonian of adatom $i=1,2$ is modelled as 
\[
H_i=\epsilon_0
\]
The hopping from graphene to adatom $i$ is expressed as a $1\times 2$ hopping matrix from sublattice space to adatom level $\epsilon_0$. It may be either 
\[
V^A_i = t\left(\begin{array}{cc} 1& 0\end{array}\right) = t\, p_A
\]
for an adatom attached to the A sublattice, or 
\[
V^B_i = t\left(\begin{array}{cc} 0& 1\end{array}\right) =  t\, p_B
\]
for the B sublattice. 

The total energy of electrons scattering on two impurities at a distance $\bm r_{12}=\bm r_2-\bm r_1$ can be expressed as
\begin{eqnarray}
U(\bm{r}_{12})&=&\int d\omega\,  (\rho(\omega)-\rho_\infty(\omega))\,\omega f(\omega)\\
\rho(\omega)&=&-\frac{1}{\pi}\mathrm{Im}\space\mathrm{Tr}\left[G^a_1(\omega)+G^a_2(\omega)+\int d^2 r G(\bm r, \bm r; \omega)\right]\nonumber
\end{eqnarray}
In this expression $f(\omega)$ is the Fermi distribution and $\rho(\omega)$ is the total density of states at energy $\omega$ of electrons in graphene (computed from the  retarded Green function $G$ of graphene) and in the two adatoms (computed from their respective $G^a_{1,2}$). The function $\rho_\infty(\omega)$ is the corresponding density of states for adatoms separated by a large distance (no interadatom scattering of electrons). 

\begin{figure}
   \centering
   \includegraphics[width=\columnwidth]{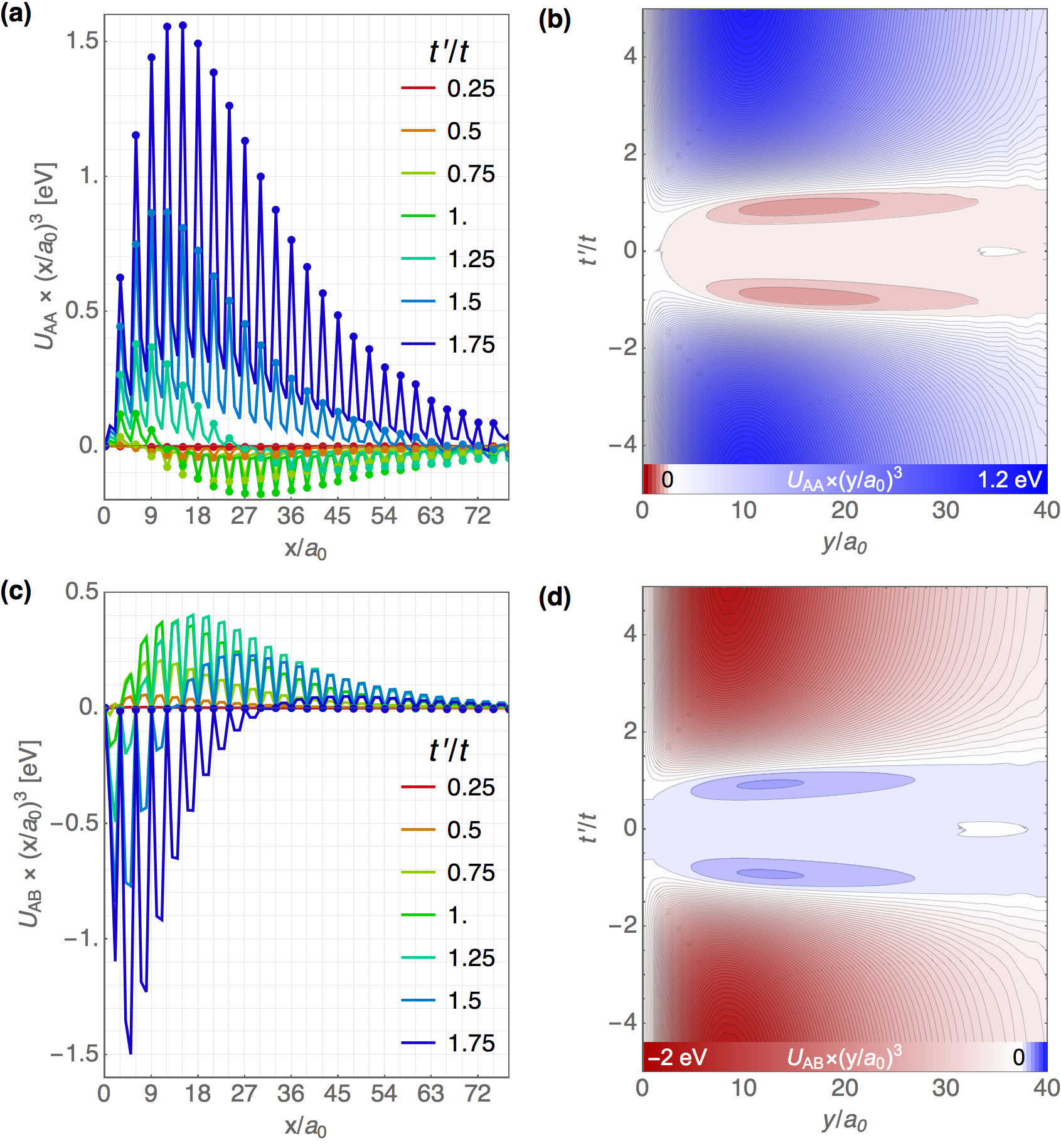}
   \caption{Interaction potential $U_{AA}$ (panels a,b) and $U_{AB}$ (panels b,c)  along the x and y direction at fixed $\epsilon_0=-0.15t$ as $t'/t$ is varied. The atomic-scale Kekulé oscillations only arise in cuts along the x direction (panels a,c). In (a,c), dots denote positions with $\nu_{\bm r}=0$ (same Kekuké character of adatoms) . $U_{AB}$ and $U_{AA}$ have an opposite sign, but the sign is inverted in an attractive-repulsive crossover that appears around $|t'|\approx 1.5 |t|$ \cite{LeBohec:PRB14}.}
   \label{fig:cuts}
\end{figure}

%%%%%%%%%%%%%%%%%%%%%%%%%%%%%%%%%%%%%%%%%%%%%%%%%%%%%%%%%%%%%%%%%%%%%%%%%%%%%%%%%%%%
\begin{figure*}
   \centering
   \includegraphics[width=1\textwidth]{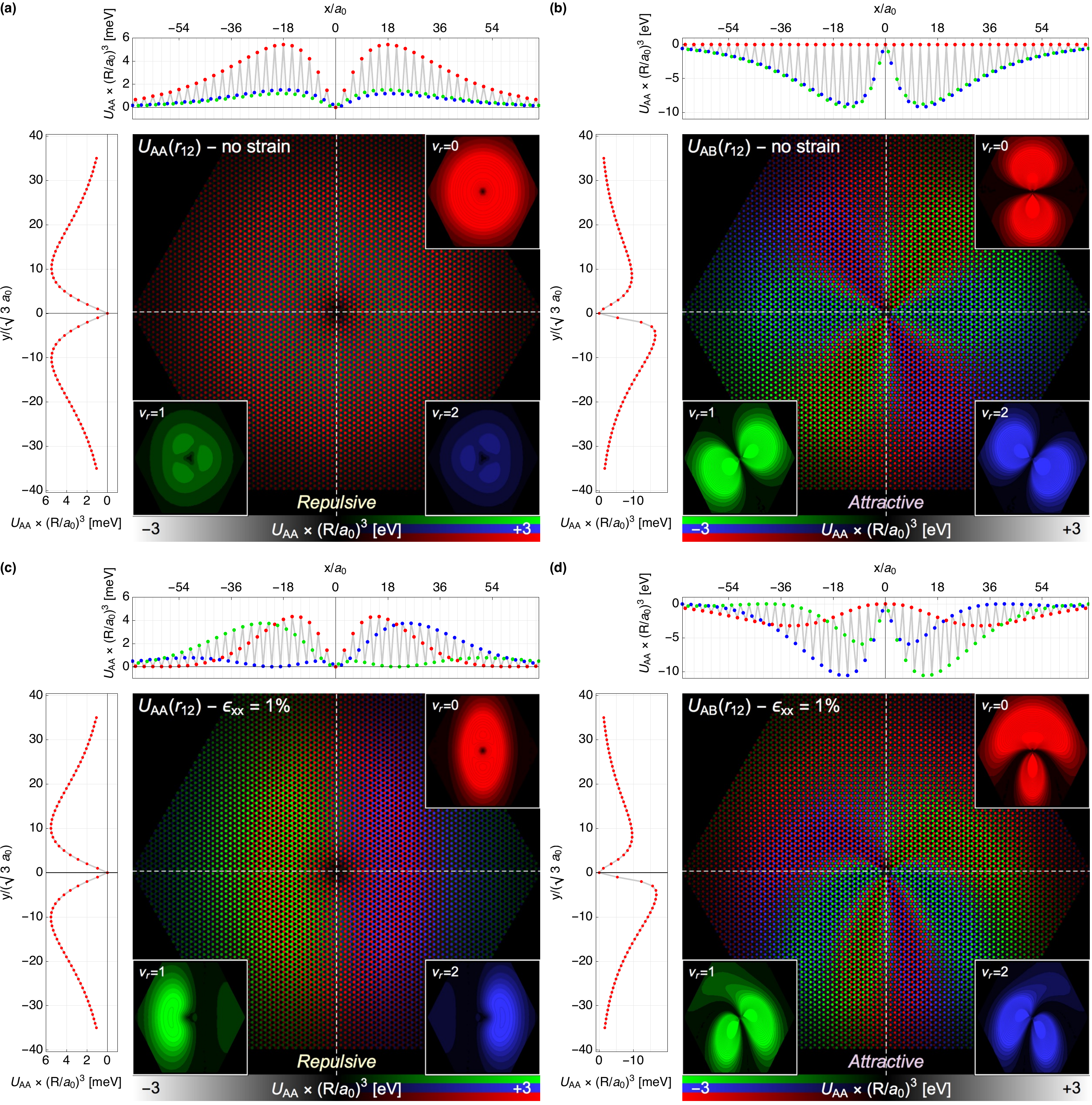}
   \caption{Spatial map of the $U_{AA}(\bm r)$ and $U_{AB}(\bm r)$ adatom interaction in the strong coupling limit ($t'=5t$, $\epsilon_0=-0.15t$). Panels (a,b) show the case without strain, and (c,d) the case with a 1\% uniaxial strain along the $x$ direction.}
   \label{fig:strong}
\end{figure*}
%%%%%%%%%%%%%%%%%%%%%%%%%%%%%%%%%%%%%%%%%%%%%%%%%%%%%%%%%%%%%%%%%%%%%%%%%%%%%%%%%%%%

The graphene Green function $G$ includes the coupling $V_i=V^{\alpha_i}_i$ of the two adatoms $i=1,2$ on sublattice $\alpha_i$, and can be derived using the Dyson equation. This yields
\begin{eqnarray}
G(\bm r, \bm r') &=& g(\bm r, \bm r') + \sum_{i,j} g(\bm r, \bm r_i)T_{ij}g(\bm r_j, \bm r')\\
g(\bm r, \bm r';\omega) &=&\sum_{s=\pm} e^{-i s\bm K \cdot(\bm r-\bm r')}\int \frac{d^2\bm k}{2\pi} e^{-i \bm k\cdot(\bm r-\bm r')}g^s(\bm k;\omega)\nonumber\\
g^s(\bm k;\omega) &=& \frac{1}{\omega-H_0(\bm k)+i0^+}
\end{eqnarray}
The $T$-matrix contains the scattering potential due to all possible inter- and intra- adatom scattering processes, and reads
\begin{eqnarray}
T_{i,j}(\omega) &=& \left(\begin{array}{cc} V^\dagger_1& 0\\0& V^\dagger_2\end{array}\right)\frac{1}{(\omega-\epsilon_0)\mathbbm{1}-\Sigma^a(\omega)}
\left(\begin{array}{cc} V_1& 0\\0& V_2\end{array}\right) \nonumber\\
\Sigma^a_{ij} &=& 
\left(\begin{array}{cc} V_1& 0\\0& V_2\end{array}\right)
\left(\begin{array}{cc} g_{11}& g_{12}\\g_{21}& g_{22}\end{array}\right)\left(\begin{array}{cc} V^\dagger_1& 0\\0& V^\dagger_2\end{array}\right),
\end{eqnarray}
where $g_{ij} = g(\bm r_i, \bm r_j; \omega)$. The expression of $T^\infty$ and $G^\infty$ for adatoms infinitely apart is obtained simply by setting $g_{12}=g_{21}=0$ above,
\begin{equation}
\Sigma^{a,\infty}_{ij} =
\left(\begin{array}{cc} V_1& 0\\0& V_2\end{array}\right)
\left(\begin{array}{cc} g_{11}& 0\\0& g_{22}\end{array}\right)\left(\begin{array}{cc} V^\dagger_1& 0\\0& V^\dagger_2\end{array}\right),
\end{equation}

%The adatom Green function on the other hand reads
%\begin{eqnarray}
%G^a_i &=& g^a_i + g^a_iT^a_ig^a_i\\
%g^a_i &=& \frac{1}{\omega-\epsilon_0 +i0^+}\\
%T^a_i &=& V_i G(\bm r_i,\bm r_i)V_i^\dagger
%\end{eqnarray}

The adatom Green function on the other hand reads
\begin{equation}
G_i = \frac{1}{\omega-\epsilon_0 - \Sigma^g_i}
\end{equation}
where the graphene-induced self-energy on adatom $i$ reads
\begin{eqnarray}
\Sigma^g_1 &=& V_1^\dagger \left(g_{11}+g_{12}t_{2}g_{21}\right)V_1\\
\Sigma^g_2 &=& V_2^\dagger \left(g_{22}+g_{21}t_{1}g_{12}\right)V_2\\
t_{j} &=& V_j \frac{1}{\omega - \epsilon_0 - \Sigma^{a,\infty}_{jj}}V_j^\dagger.
\end{eqnarray}
The asymptotic $G_i^\infty$ at large adatom separation is obtained by setting $g_{12}=g_{21}=0$ above.

With these ingredients, the final form for the density of states reads
\begin{eqnarray}
\rho(\omega)&-&\rho_\infty(\omega)= \Delta\rho^g(\omega)+\Delta\rho^1(\omega)+\Delta\rho^2(\omega) \\
\Delta\rho^g &=& -\frac{1}{\pi}\mathrm{Im}\,\mathrm{Tr}\sum_{s,i,j}\int\frac{d^2k}{2\pi}g^s(\bm k)\Delta T_{ij}g^s(\bm k)e^{-i\bm k\cdot(\bm r_j-\bm r_i)}\nonumber\\
\Delta\rho^i &=& -\frac{1}{\pi}\mathrm{Im}\,\mathrm{Tr}\,\Delta G_i
\end{eqnarray}
where $\Delta T_{ij}=T_{ij}-T^\infty_{ij}$ and $\Delta G_i=G_i-G_i^\infty$. Alternative derivations with different (but formally equivalent) forms of these equations can be found e.g. in Refs. \cite{Hyldgaard:JOPCM00,Shytov:PRL09,LeBohec:PRB14}.

%%%%%%%%%%%%%%%%%%%%%%%%%%%%%%%%%%%%%%%%%%%%%%%%%%%%%%%%%%%%%%%%%%%%%%%%%%%%%%%%%%%%
\begin{figure}
   \centering
   \includegraphics[width=\columnwidth]{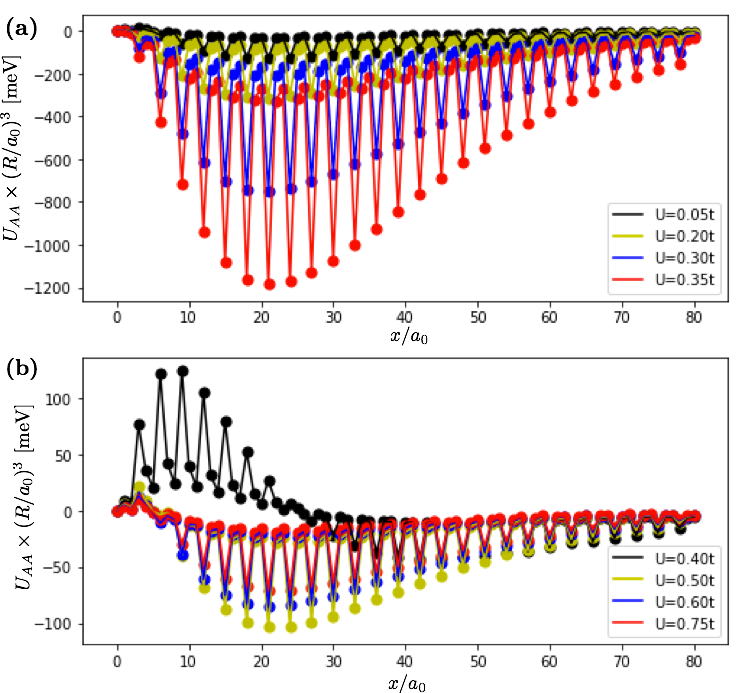}
   \caption{Interaction potential for same-sublattice adatoms along the $x$ axis in the presence of an onsite electron-electron repulsion U. Panel (a) shows the interaction potential for values of $U$ below $U_c$ . Panel (b) shows the evolution of the interaction potential in the ferromagnetic regime $U> U_c$.   }
   \label{fig:cuts_vs_U}
\end{figure}
%%%%%%%%%%%%%%%%%%%%%%%%%%%%%%%%%%%%%%%%%%%%%%%%%%%%%%%%%%%%%%%%%%%%%%%%%%%%%%%%%%%% 

\begin{figure}
   \centering
   \includegraphics[width=0.9\columnwidth]{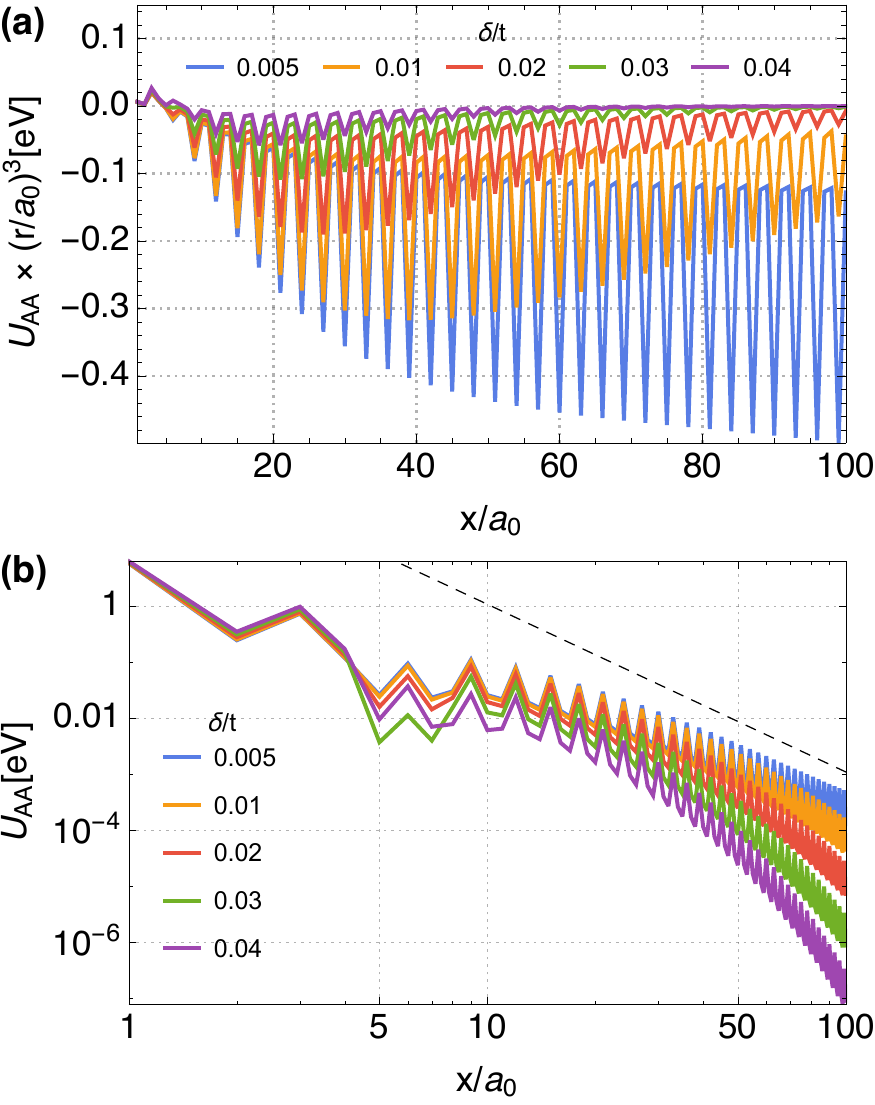}
   \caption{$U_{AA}(r)$ cut at $y=0$ analogous to Fig. 2a of the main text as the damping factor of electrons in graphene is reduced from $0.04t$ to $0.005t$. The value chosen in the main text is $\delta=0.03t$. Note that as coherence increases, the potential becomes stronger and decays as $1/r^3$ (dashed line) at large distances.}
   \label{fig:rcube}
\end{figure}

\section{Strongly coupled adatoms}
\label{ap:strong}

In the main text, we have concentrated on the weak coupling regime $t'/t < 1$ in which the Kekulé interaction potential is attractive when the adatoms lie on the same-sublattice, and repulsive otherwise. This type of interaction has been observed experimentally at room temperatures for a specific type of `vacancy adatom' \cite{Gutierrez:NP16}.  The magnitude and even the sign of the interaction, however, strongly depends on the ratio $t'/t$, i.e. on how strongly the adatom or atomic defect binds to graphene. In this section we explore the dependence of the Kekulé interaction as the coupling $t'$ is increased. We find that the strong-coupling regime $t'/t \gg 1$ is characterised by a repulsive Kekulé potential for same-sublattice configurations and an attractive potential for opposite sublattices, as noted by previous works \cite{LeBohec:PRB14}. The boundary between the weak-coupling and strong-coupling regimes is found at $t' \approx 1.5t$. This is shown in
Fig. \ref{fig:cuts}, where we plot the dependence of the $U_{AA}$ and $U_{AB}$ potentials along the $x$ and $y$ direction as one increases the ratio $t'/t$ for a fixed $\epsilon_0\approx -0.15 t$.

Figure \ref{fig:strong} shows the corresponding map for all interadatom distances $\bm{r}$ in the strong coupling limit $t'=5t$. In panels (a,b) we show the case without any strain. We see that, indeed, same-sublattice interaction is now repulsive (a), and different sublattice interaction is attractive (b). Sublattice ordering will thus tend to be `antiferromagnetic', with nearby adatoms arranging in opposite lattices. Without strain, different Kekulé alignments are favored (panel b) depending on the angle $\phi$ between $\bm r$ and $\Delta \bm K$ (here along the x direction). The $U_{AB}$ potential that dominates the arrangement of adatoms is therefore non-isotropic, in contrast to the $U_{AA}$ potential that controls the weak coupling regime. Most importantly, the magnitude of the adatom interaction is between one and two orders of magnitude stronger than in the weak coupling regime.

In the presence of strain, the interaction potential becomes modulated following the same pseudogauge mechanism described in the main text. However, since adatom ordering in the strong coupling regime is controlled by the non-istropic potential $U_{AB}$, the effect of strain has a much richer structure in this case, see Figure \ref{fig:strong}d.

%\begin{figure}
%   \centering
%   \includegraphics[width=0.4\textwidth]{energy_vs_t'.png}
%   \caption{Interaction potential along the x direction for same-sublattice (panel a) and opposite-sublattice (panel b) for a set of values of $t'$.}
%   \label{fig:strong}
%\end{figure}

%%%%%%%%%%%%%%%%%%%%%%%%%%%%%%%%%%%%%%%%%%%%%%%%%%%%%%%%%%%%%%%%%%%%%%%%%%%%%%%%%%%%%
%\begin{figure*}
%   \centering
%   \includegraphics[width=\textwidth]{strong-hor.pdf}
%   \caption{example caption}
%   \label{fig:strong}
%\end{figure*}
%%%%%%%%%%%%%%%%%%%%%%%%%%%%%%%%%%%%%%%%%%%%%%%%%%%%%%%%%%%%%%%%%%%%%%%%%%%%%%%%%%%%%

%%%%%%%%%%%%%%%%%%%%%%%%%%%%%%%%%%%%%%%%%%%%%%%%%%%%%%%%%%%%%%%%%%%%%%%%%%%%%%%%%%%%%
%\begin{figure}
%   \centering
%   \includegraphics[height=\textheight]{strong-ver.pdf}
%   \caption{example caption}
%   \label{fig:strong}
%\end{figure}
%%%%%%%%%%%%%%%%%%%%%%%%%%%%%%%%%%%%%%%%%%%%%%%%%%%%%%%%%%%%%%%%%%%%%%%%%%%%%%%%%%%%%

\section{Interactions}
\label{ap:interactions}

Thus far, we have not considered the effects of electron-electron interactions in our discussion. In this section we consider intra-adatom Hubbard interactions in the weak coupling limit. The Hubbard interaction introduces an additional term in the adatom Hamiltonian. In the mean-field approximation it is expressed as: 
\begin{equation}\label{H_new}
H_{j} = \epsilon_{0} + U\sum_{\sigma,\sigma^{\prime}} \langle n_{j,\sigma} \rangle n_{j,\sigma^{\prime}}
\end{equation}  
where $U$ is the intensity of the Hubbard interaction, $n_{j\sigma}$ is the number operator for an electron in adatom $j=1,2$ with spin $\sigma = \uparrow,\downarrow$ and the magnetic moment in the adatom is $M_{j} = \langle n_{j,\uparrow}\rangle - \langle n_{j,\downarrow}\rangle$ is to be computed self-consistently. 

Assume two adatoms $j=1,2$ on graphene. The mean value of the number of electrons with spin-label $\sigma$ in adatom $j$ reads:
\begin{equation}
\langle n_{j\sigma} \rangle = -\frac{1}{\pi} \mathrm{Im} \int d\omega \left( \omega+i0^{+}-H_{j} -\Sigma^{gr,j^{\prime}} (\omega)    \right) \omega
\end{equation}  
and $\Sigma^{gr,j^{\prime}} (\omega)  = -t^{\prime\dagger}G(\bm{r_{j}},\bm{r_{j}}) t^{\prime}$ is the self-energy term accounting for the combined influence of the graphene lattice and the $j^{\prime}\neq j$ adatom on the $j$ adatom. The Green function $G(\bm{r_{j}},\bm{r_{j}})$ of graphene under the $j$ adatom includes the presence of the $j^{\prime}$ adatom. It is computed by the same procedure explained in Appendix \ref{ap:method} for the non-interacting case, with the modification that the adatom Hamiltonian $H_j$ is now given by equation (\ref{H_new}) and is computed self-consistently. The formula for the potential $U$ given in the main text holds unmodified.

The spin-exchange interaction between magnetic adatoms in top positions is ferromagnetic when the adatoms are located in the same sublattice and antiferromagnetic when located in opposite sublattices \cite{Sherafati:PRB11,Black-Schaffer:PRB10,Gonzalez-Herrero:S16}. We have confirmed this result within our model, and have checked that the ferromagnetic character of the exchange remains unchanged under the application of strain in graphene. In the unstrained case, in the ferromagnetic regime the Kekule $\cos^2 \left( \Delta\bm{K\cdot r}\right)$ periodicity is left intact. Only the envelope $v_{AA}$ and $v_{AB}$ of the oscillations is modified by the effect of $U$.

If the adatom is decoupled from graphene ($t^{\prime} = 0$), the presence of an arbitrarily small $U$ would open a spin-polarized splitting in the low-energy spectrum of the adatom.  For $t^{\prime} \neq 0$, our mean field approximation gives a minimum $U_c>0$ required to create a non-zero magnetic moment in the adatoms. For $t^{\prime} = 0.7 t$ and $\epsilon=-0.15t$ , $U_c \approx 0.4t$. Our numerical calculations show that the effect of the electron-electron repulsion is two-fold. For $U<U_c$, the depth of the potential well increases with $U$, thus enhancing the attractive strength of the Kekulé ordering. In the regime of ferromagnetic alignment ($U> U_c$), the effect on the envelope is somewhat more complicated. For $U$ very close to $U_c$ the repulsive core  around $\bm{r }= 0$ is increased, although the interaction quickly becomes attractive for longer distances. Upon further increase of $U$ the repulsive core shrinks dramatically and the system returns to a behavior similar to the non-magnetic case. This behavior can be observed in Fig. \ref{fig:cuts_vs_U}.

\section{Dissipation and the $1/r^3$ asymptotics}
\label{ap:dissipation}

It may be shown analytically \cite{LeBohec:PRB14} that a pristine and fully coherent graphene substrate leads to a same-sublattice adatom potential $U_{AA}(\bm r)=v_{AA}(r)\cos(2\pi\nu_{\bf{r}})$ that scales asymptotically as $v_{AA}(r)\sim 1/r^3$ with interadatom distance (at shorter distances, deviations are predicted depending on the adatom coupling strength \cite{Shytov:PRL09}). This asymptotic result, however, assumes that dissipation is completely absent in the graphene electron liquid. Inelastic scattering events with phonons or through electron-electron interactions modify this result. In the main text, our simulations incorporated phenomenologically electronic dissipation by a finite imaginary part $\delta=0.03t$ added to the energy $\omega+i\delta$ in the bare Green's functions $g$. The precise value of $\delta$ adequate for a real system is model-dependent. Its effect on $v_{AA}$, however, is quite universal, and leads to a suppression of the interaction strength and a faster decay than $1/r^3$ at long distances. To make connection to the analytical results for fully coherent systems we present in this section results for $U_{AA}(\bm r)$ as the damping factor $\delta$ is reduced. Fig. \ref{fig:rcube} shows cuts at $y=0$ analogous to those in Fig. 2a in the main text as $\delta$ is reduced from $0.04t$ to $0.005t$, both in an $U_{AA}(r)\times (r/a_0)^3$ plot (panel a) as in a log-log plot (panel b). We see clearly that the interaction strength is enhanced as the system becomes more coherent, and that the $1/r^3$ decay (dashed line in panel b) is recovered.

\bibliography{/Users/pablo/Dropbox/Bibliography/BibdeskPablo/biblio}

\begin{thebibliography}{41}%
\makeatletter
\providecommand \@ifxundefined [1]{%
 \@ifx{#1\undefined}
}%
\providecommand \@ifnum [1]{%
 \ifnum #1\expandafter \@firstoftwo
 \else \expandafter \@secondoftwo
 \fi
}%
\providecommand \@ifx [1]{%
 \ifx #1\expandafter \@firstoftwo
 \else \expandafter \@secondoftwo
 \fi
}%
\providecommand \natexlab [1]{#1}%
\providecommand \enquote  [1]{``#1''}%
\providecommand \bibnamefont  [1]{#1}%
\providecommand \bibfnamefont [1]{#1}%
\providecommand \citenamefont [1]{#1}%
\providecommand \href@noop [0]{\@secondoftwo}%
\providecommand \href [0]{\begingroup \@sanitize@url \@href}%
\providecommand \@href[1]{\@@startlink{#1}\@@href}%
\providecommand \@@href[1]{\endgroup#1\@@endlink}%
\providecommand \@sanitize@url [0]{\catcode `\\12\catcode `\$12\catcode
  `\&12\catcode `\#12\catcode `\^12\catcode `\_12\catcode `\%12\relax}%
\providecommand \@@startlink[1]{}%
\providecommand \@@endlink[0]{}%
\providecommand \url  [0]{\begingroup\@sanitize@url \@url }%
\providecommand \@url [1]{\endgroup\@href {#1}{\urlprefix }}%
\providecommand \urlprefix  [0]{URL }%
\providecommand \Eprint [0]{\href }%
\providecommand \doibase [0]{http://dx.doi.org/}%
\providecommand \selectlanguage [0]{\@gobble}%
\providecommand \bibinfo  [0]{\@secondoftwo}%
\providecommand \bibfield  [0]{\@secondoftwo}%
\providecommand \translation [1]{[#1]}%
\providecommand \BibitemOpen [0]{}%
\providecommand \bibitemStop [0]{}%
\providecommand \bibitemNoStop [0]{.\EOS\space}%
\providecommand \EOS [0]{\spacefactor3000\relax}%
\providecommand \BibitemShut  [1]{\csname bibitem#1\endcsname}%
\let\auto@bib@innerbib\@empty
%</preamble>
\bibitem [{\citenamefont {Neto}\ \emph {et~al.}(2009)\citenamefont {Neto},
  \citenamefont {Guinea}, \citenamefont {Peres}, \citenamefont {Novoselov},\
  and\ \citenamefont {Geim}}]{Neto:RMP09}%
  \BibitemOpen
  \bibfield  {author} {\bibinfo {author} {\bibfnamefont {A.~H.~C.}\
  \bibnamefont {Neto}}, \bibinfo {author} {\bibfnamefont {F.}~\bibnamefont
  {Guinea}}, \bibinfo {author} {\bibfnamefont {N.~M.~R.}\ \bibnamefont
  {Peres}}, \bibinfo {author} {\bibfnamefont {K.~S.}\ \bibnamefont
  {Novoselov}}, \ and\ \bibinfo {author} {\bibfnamefont {A.~K.}\ \bibnamefont
  {Geim}},\ }\href@noop {} {\bibfield  {journal} {\bibinfo  {journal} {Rev.
  Mod. Phys.}\ }\textbf {\bibinfo {volume} {81}},\ \bibinfo {pages} {109}
  (\bibinfo {year} {2009})}\BibitemShut {NoStop}%
\bibitem [{\citenamefont {Das~Sarma}\ \emph {et~al.}(2011)\citenamefont
  {Das~Sarma}, \citenamefont {Adam}, \citenamefont {Hwang},\ and\ \citenamefont
  {Rossi}}]{Das-Sarma:RMP11}%
  \BibitemOpen
  \bibfield  {author} {\bibinfo {author} {\bibfnamefont {S.}~\bibnamefont
  {Das~Sarma}}, \bibinfo {author} {\bibfnamefont {S.}~\bibnamefont {Adam}},
  \bibinfo {author} {\bibfnamefont {E.~H.}\ \bibnamefont {Hwang}}, \ and\
  \bibinfo {author} {\bibfnamefont {E.}~\bibnamefont {Rossi}},\ }\href
  {\doibase 10.1103/RevModPhys.83.407} {\bibfield  {journal} {\bibinfo
  {journal} {Rev. Mod. Phys.}\ }\textbf {\bibinfo {volume} {83}},\ \bibinfo
  {pages} {407} (\bibinfo {year} {2011})}\BibitemShut {NoStop}%
\bibitem [{\citenamefont {Amorim}\ \emph {et~al.}(2016)\citenamefont {Amorim},
  \citenamefont {Cortijo}, \citenamefont {de~Juan}, \citenamefont {Grushin},
  \citenamefont {Guinea}, \citenamefont {Guti{\'e}rrez-Rubio}, \citenamefont
  {Ochoa}, \citenamefont {Parente}, \citenamefont {Rold{\'a}n}, \citenamefont
  {San-Jose}, \citenamefont {Schiefele}, \citenamefont {Sturla},\ and\
  \citenamefont {Vozmediano}}]{Amorim:PR16}%
  \BibitemOpen
  \bibfield  {author} {\bibinfo {author} {\bibfnamefont {B.}~\bibnamefont
  {Amorim}}, \bibinfo {author} {\bibfnamefont {A.}~\bibnamefont {Cortijo}},
  \bibinfo {author} {\bibfnamefont {F.}~\bibnamefont {de~Juan}}, \bibinfo
  {author} {\bibfnamefont {A.}~\bibnamefont {Grushin}}, \bibinfo {author}
  {\bibfnamefont {F.}~\bibnamefont {Guinea}}, \bibinfo {author} {\bibfnamefont
  {A.}~\bibnamefont {Guti{\'e}rrez-Rubio}}, \bibinfo {author} {\bibfnamefont
  {H.}~\bibnamefont {Ochoa}}, \bibinfo {author} {\bibfnamefont
  {V.}~\bibnamefont {Parente}}, \bibinfo {author} {\bibfnamefont
  {R.}~\bibnamefont {Rold{\'a}n}}, \bibinfo {author} {\bibfnamefont
  {P.}~\bibnamefont {San-Jose}}, \bibinfo {author} {\bibfnamefont
  {J.}~\bibnamefont {Schiefele}}, \bibinfo {author} {\bibfnamefont
  {M.}~\bibnamefont {Sturla}}, \ and\ \bibinfo {author} {\bibfnamefont
  {M.}~\bibnamefont {Vozmediano}},\ }\href {\doibase
  http://dx.doi.org/10.1016/j.physrep.2015.12.006} {\bibfield  {journal}
  {\bibinfo  {journal} {Physics Reports}\ }\textbf {\bibinfo {volume} {617}},\
  \bibinfo {pages} {1 } (\bibinfo {year} {2016})},\ \bibinfo {note} {novel
  effects of strains in graphene and other two dimensional
  materials}\BibitemShut {NoStop}%
\bibitem [{\citenamefont {Bolotin}\ \emph {et~al.}(2008)\citenamefont
  {Bolotin}, \citenamefont {Sikes}, \citenamefont {Jiang}, \citenamefont
  {Klima}, \citenamefont {Fudenberg}, \citenamefont {Hone}, \citenamefont
  {Kim},\ and\ \citenamefont {Stormer}}]{Bolotin:SSC08}%
  \BibitemOpen
  \bibfield  {author} {\bibinfo {author} {\bibfnamefont {K.}~\bibnamefont
  {Bolotin}}, \bibinfo {author} {\bibfnamefont {K.}~\bibnamefont {Sikes}},
  \bibinfo {author} {\bibfnamefont {Z.}~\bibnamefont {Jiang}}, \bibinfo
  {author} {\bibfnamefont {M.}~\bibnamefont {Klima}}, \bibinfo {author}
  {\bibfnamefont {G.}~\bibnamefont {Fudenberg}}, \bibinfo {author}
  {\bibfnamefont {J.}~\bibnamefont {Hone}}, \bibinfo {author} {\bibfnamefont
  {P.}~\bibnamefont {Kim}}, \ and\ \bibinfo {author} {\bibfnamefont
  {H.}~\bibnamefont {Stormer}},\ }\href {\doibase DOI:
  10.1016/j.ssc.2008.02.024} {\bibfield  {journal} {\bibinfo  {journal} {Solid
  State Commun.}\ }\textbf {\bibinfo {volume} {146}},\ \bibinfo {pages} {351 }
  (\bibinfo {year} {2008})}\BibitemShut {NoStop}%
\bibitem [{\citenamefont {Beenakker}(2008)}]{Beenakker:RMP08}%
  \BibitemOpen
  \bibfield  {author} {\bibinfo {author} {\bibfnamefont {C.}~\bibnamefont
  {Beenakker}},\ }\href@noop {} {\bibfield  {journal} {\bibinfo  {journal}
  {Rev. Mod. Phys.}\ }\textbf {\bibinfo {volume} {80}},\ \bibinfo {pages}
  {1337} (\bibinfo {year} {2008})}\BibitemShut {NoStop}%
\bibitem [{\citenamefont {Gorbachev}\ \emph {et~al.}(2014)\citenamefont
  {Gorbachev}, \citenamefont {Song}, \citenamefont {Yu}, \citenamefont
  {Kretinin}, \citenamefont {Withers}, \citenamefont {Cao}, \citenamefont
  {Mishchenko}, \citenamefont {Grigorieva}, \citenamefont {Novoselov},
  \citenamefont {Levitov},\ and\ \citenamefont {Geim}}]{Gorbachev:S14}%
  \BibitemOpen
  \bibfield  {author} {\bibinfo {author} {\bibfnamefont {R.~V.}\ \bibnamefont
  {Gorbachev}}, \bibinfo {author} {\bibfnamefont {J.~C.~W.}\ \bibnamefont
  {Song}}, \bibinfo {author} {\bibfnamefont {G.~L.}\ \bibnamefont {Yu}},
  \bibinfo {author} {\bibfnamefont {A.~V.}\ \bibnamefont {Kretinin}}, \bibinfo
  {author} {\bibfnamefont {F.}~\bibnamefont {Withers}}, \bibinfo {author}
  {\bibfnamefont {Y.}~\bibnamefont {Cao}}, \bibinfo {author} {\bibfnamefont
  {A.}~\bibnamefont {Mishchenko}}, \bibinfo {author} {\bibfnamefont {I.~V.}\
  \bibnamefont {Grigorieva}}, \bibinfo {author} {\bibfnamefont {K.~S.}\
  \bibnamefont {Novoselov}}, \bibinfo {author} {\bibfnamefont {L.~S.}\
  \bibnamefont {Levitov}}, \ and\ \bibinfo {author} {\bibfnamefont {A.~K.}\
  \bibnamefont {Geim}},\ }\href {\doibase 10.1126/science.1254966} {\bibfield
  {journal} {\bibinfo  {journal} {Science}\ }\textbf {\bibinfo {volume}
  {346}},\ \bibinfo {pages} {448} (\bibinfo {year} {2014})}\BibitemShut
  {NoStop}%
\bibitem [{\citenamefont {Martin}\ \emph {et~al.}(2008)\citenamefont {Martin},
  \citenamefont {Blanter},\ and\ \citenamefont {Morpurgo}}]{Martin:PRL08}%
  \BibitemOpen
  \bibfield  {author} {\bibinfo {author} {\bibfnamefont {I.}~\bibnamefont
  {Martin}}, \bibinfo {author} {\bibfnamefont {Y.~M.}\ \bibnamefont {Blanter}},
  \ and\ \bibinfo {author} {\bibfnamefont {A.~F.}\ \bibnamefont {Morpurgo}},\
  }\href@noop {} {\bibfield  {journal} {\bibinfo  {journal} {Phys. Rev. Lett.}\
  }\textbf {\bibinfo {volume} {100}},\ \bibinfo {pages} {036804} (\bibinfo
  {year} {2008})}\BibitemShut {NoStop}%
\bibitem [{\citenamefont {San-Jose}\ and\ \citenamefont
  {Prada}(2013)}]{San-Jose:PRB13}%
  \BibitemOpen
  \bibfield  {author} {\bibinfo {author} {\bibfnamefont {P.}~\bibnamefont
  {San-Jose}}\ and\ \bibinfo {author} {\bibfnamefont {E.}~\bibnamefont
  {Prada}},\ }\href {\doibase 10.1103/PhysRevB.88.121408} {\bibfield  {journal}
  {\bibinfo  {journal} {Phys. Rev. B}\ }\textbf {\bibinfo {volume} {88}},\
  \bibinfo {pages} {121408} (\bibinfo {year} {2013})}\BibitemShut {NoStop}%
\bibitem [{\citenamefont {Chen}\ \emph {et~al.}(2009)\citenamefont {Chen},
  \citenamefont {Cullen}, \citenamefont {Jang}, \citenamefont {Fuhrer},\ and\
  \citenamefont {Williams}}]{Chen:PRL09}%
  \BibitemOpen
  \bibfield  {author} {\bibinfo {author} {\bibfnamefont {J.-H.}\ \bibnamefont
  {Chen}}, \bibinfo {author} {\bibfnamefont {W.~G.}\ \bibnamefont {Cullen}},
  \bibinfo {author} {\bibfnamefont {C.}~\bibnamefont {Jang}}, \bibinfo {author}
  {\bibfnamefont {M.~S.}\ \bibnamefont {Fuhrer}}, \ and\ \bibinfo {author}
  {\bibfnamefont {E.~D.}\ \bibnamefont {Williams}},\ }\href {\doibase
  10.1103/PhysRevLett.102.236805} {\bibfield  {journal} {\bibinfo  {journal}
  {Phys. Rev. Lett.}\ }\textbf {\bibinfo {volume} {102}},\ \bibinfo {pages}
  {236805} (\bibinfo {year} {2009})}\BibitemShut {NoStop}%
\bibitem [{\citenamefont {Cresti}\ and\ \citenamefont
  {Roche}(2009)}]{Cresti:PRB09}%
  \BibitemOpen
  \bibfield  {author} {\bibinfo {author} {\bibfnamefont {A.}~\bibnamefont
  {Cresti}}\ and\ \bibinfo {author} {\bibfnamefont {S.}~\bibnamefont {Roche}},\
  }\href {\doibase 10.1103/PhysRevB.79.233404} {\bibfield  {journal} {\bibinfo
  {journal} {Phys. Rev. B}\ }\textbf {\bibinfo {volume} {79}},\ \bibinfo
  {pages} {233404} (\bibinfo {year} {2009})}\BibitemShut {NoStop}%
\bibitem [{\citenamefont {Libisch}\ \emph {et~al.}(2012)\citenamefont
  {Libisch}, \citenamefont {Rotter},\ and\ \citenamefont
  {Burgd{\"o}rfer}}]{Libisch:NJP12}%
  \BibitemOpen
  \bibfield  {author} {\bibinfo {author} {\bibfnamefont {F.}~\bibnamefont
  {Libisch}}, \bibinfo {author} {\bibfnamefont {S.}~\bibnamefont {Rotter}}, \
  and\ \bibinfo {author} {\bibfnamefont {J.}~\bibnamefont {Burgd{\"o}rfer}},\
  }\href {\doibase 10.1088/1367-2630/14/12/123006} {\bibfield  {journal}
  {\bibinfo  {journal} {New J. Phys.}\ }\textbf {\bibinfo {volume} {14}},\
  \bibinfo {pages} {123006} (\bibinfo {year} {2012})}\BibitemShut {NoStop}%
\bibitem [{\citenamefont {Lawlor}\ \emph {et~al.}(2013)\citenamefont {Lawlor},
  \citenamefont {Power},\ and\ \citenamefont {Ferreira}}]{Lawlor:PRB13}%
  \BibitemOpen
  \bibfield  {author} {\bibinfo {author} {\bibfnamefont {J.~A.}\ \bibnamefont
  {Lawlor}}, \bibinfo {author} {\bibfnamefont {S.~R.}\ \bibnamefont {Power}}, \
  and\ \bibinfo {author} {\bibfnamefont {M.~S.}\ \bibnamefont {Ferreira}},\
  }\href {\doibase 10.1103/PhysRevB.88.205416} {\bibfield  {journal} {\bibinfo
  {journal} {Phys. Rev. B}\ }\textbf {\bibinfo {volume} {88}},\ \bibinfo
  {pages} {205416} (\bibinfo {year} {2013})}\BibitemShut {NoStop}%
\bibitem [{\citenamefont {Zhao}\ \emph {et~al.}(2011)\citenamefont {Zhao},
  \citenamefont {He}, \citenamefont {Rim}, \citenamefont {Schiros},
  \citenamefont {Kim}, \citenamefont {Zhou}, \citenamefont {Guti{\'e}rrez},
  \citenamefont {Chockalingam}, \citenamefont {Arguello}, \citenamefont
  {P{\'a}lov{\'a}}, \citenamefont {Nordlund}, \citenamefont {Hybertsen},
  \citenamefont {Reichman}, \citenamefont {Heinz}, \citenamefont {Kim},
  \citenamefont {Pinczuk}, \citenamefont {Flynn},\ and\ \citenamefont
  {Pasupathy}}]{Zhao:S11}%
  \BibitemOpen
  \bibfield  {author} {\bibinfo {author} {\bibfnamefont {L.}~\bibnamefont
  {Zhao}}, \bibinfo {author} {\bibfnamefont {R.}~\bibnamefont {He}}, \bibinfo
  {author} {\bibfnamefont {K.~T.}\ \bibnamefont {Rim}}, \bibinfo {author}
  {\bibfnamefont {T.}~\bibnamefont {Schiros}}, \bibinfo {author} {\bibfnamefont
  {K.~S.}\ \bibnamefont {Kim}}, \bibinfo {author} {\bibfnamefont
  {H.}~\bibnamefont {Zhou}}, \bibinfo {author} {\bibfnamefont {C.}~\bibnamefont
  {Guti{\'e}rrez}}, \bibinfo {author} {\bibfnamefont {S.~P.}\ \bibnamefont
  {Chockalingam}}, \bibinfo {author} {\bibfnamefont {C.~J.}\ \bibnamefont
  {Arguello}}, \bibinfo {author} {\bibfnamefont {L.}~\bibnamefont
  {P{\'a}lov{\'a}}}, \bibinfo {author} {\bibfnamefont {D.}~\bibnamefont
  {Nordlund}}, \bibinfo {author} {\bibfnamefont {M.~S.}\ \bibnamefont
  {Hybertsen}}, \bibinfo {author} {\bibfnamefont {D.~R.}\ \bibnamefont
  {Reichman}}, \bibinfo {author} {\bibfnamefont {T.~F.}\ \bibnamefont {Heinz}},
  \bibinfo {author} {\bibfnamefont {P.}~\bibnamefont {Kim}}, \bibinfo {author}
  {\bibfnamefont {A.}~\bibnamefont {Pinczuk}}, \bibinfo {author} {\bibfnamefont
  {G.~W.}\ \bibnamefont {Flynn}}, \ and\ \bibinfo {author} {\bibfnamefont
  {A.~N.}\ \bibnamefont {Pasupathy}},\ }\href {\doibase
  10.1126/science.1208759} {\bibfield  {journal} {\bibinfo  {journal}
  {Science}\ }\textbf {\bibinfo {volume} {333}},\ \bibinfo {pages} {999}
  (\bibinfo {year} {2011})},\ \Eprint
  {http://arxiv.org/abs/http://science.sciencemag.org/content/333/6045/999.full.pdf}
  {http://science.sciencemag.org/content/333/6045/999.full.pdf} \BibitemShut
  {NoStop}%
\bibitem [{\citenamefont {Pa{\v s}ti}\ \emph {et~al.}(2017)\citenamefont {Pa{\v
  s}ti}, \citenamefont {Jovanovi{\'c}}, \citenamefont {Dobrota}, \citenamefont
  {Mentus}, \citenamefont {Johansson},\ and\ \citenamefont
  {Skorodumova}}]{Pasti:17}%
  \BibitemOpen
  \bibfield  {author} {\bibinfo {author} {\bibfnamefont {I.~A.}\ \bibnamefont
  {Pa{\v s}ti}}, \bibinfo {author} {\bibfnamefont {A.}~\bibnamefont
  {Jovanovi{\'c}}}, \bibinfo {author} {\bibfnamefont {A.~S.}\ \bibnamefont
  {Dobrota}}, \bibinfo {author} {\bibfnamefont {S.~V.}\ \bibnamefont {Mentus}},
  \bibinfo {author} {\bibfnamefont {B.}~\bibnamefont {Johansson}}, \ and\
  \bibinfo {author} {\bibfnamefont {N.~V.}\ \bibnamefont {Skorodumova}},\
  }\href {https://arxiv.org/abs/1710.10084} {\  (\bibinfo {year} {2017})},\
  \Eprint {http://arxiv.org/abs/1710.10084} {1710.10084} \BibitemShut {NoStop}%
\bibitem [{\citenamefont {Pachoud}\ \emph {et~al.}(2014)\citenamefont
  {Pachoud}, \citenamefont {Ferreira}, \citenamefont {\"Ozyilmaz},\ and\
  \citenamefont {Castro~Neto}}]{Pachoud:PRB14}%
  \BibitemOpen
  \bibfield  {author} {\bibinfo {author} {\bibfnamefont {A.}~\bibnamefont
  {Pachoud}}, \bibinfo {author} {\bibfnamefont {A.}~\bibnamefont {Ferreira}},
  \bibinfo {author} {\bibfnamefont {B.}~\bibnamefont {\"Ozyilmaz}}, \ and\
  \bibinfo {author} {\bibfnamefont {A.~H.}\ \bibnamefont {Castro~Neto}},\
  }\href {\doibase 10.1103/PhysRevB.90.035444} {\bibfield  {journal} {\bibinfo
  {journal} {Phys. Rev. B}\ }\textbf {\bibinfo {volume} {90}},\ \bibinfo
  {pages} {035444} (\bibinfo {year} {2014})}\BibitemShut {NoStop}%
\bibitem [{\citenamefont {Nair}\ \emph {et~al.}(2010)\citenamefont {Nair},
  \citenamefont {Ren}, \citenamefont {Jalil}, \citenamefont {Riaz},
  \citenamefont {Kravets}, \citenamefont {Britnell}, \citenamefont {Blake},
  \citenamefont {Schedin}, \citenamefont {Mayorov}, \citenamefont {Yuan} \emph
  {et~al.}}]{Nair:S10}%
  \BibitemOpen
  \bibfield  {author} {\bibinfo {author} {\bibfnamefont {R.~R.}\ \bibnamefont
  {Nair}}, \bibinfo {author} {\bibfnamefont {W.}~\bibnamefont {Ren}}, \bibinfo
  {author} {\bibfnamefont {R.}~\bibnamefont {Jalil}}, \bibinfo {author}
  {\bibfnamefont {I.}~\bibnamefont {Riaz}}, \bibinfo {author} {\bibfnamefont
  {V.~G.}\ \bibnamefont {Kravets}}, \bibinfo {author} {\bibfnamefont
  {L.}~\bibnamefont {Britnell}}, \bibinfo {author} {\bibfnamefont
  {P.}~\bibnamefont {Blake}}, \bibinfo {author} {\bibfnamefont
  {F.}~\bibnamefont {Schedin}}, \bibinfo {author} {\bibfnamefont {A.~S.}\
  \bibnamefont {Mayorov}}, \bibinfo {author} {\bibfnamefont {S.}~\bibnamefont
  {Yuan}},  \emph {et~al.},\ }\href@noop {} {\bibfield  {journal} {\bibinfo
  {journal} {Small}\ }\textbf {\bibinfo {volume} {6}},\ \bibinfo {pages} {2877}
  (\bibinfo {year} {2010})}\BibitemShut {NoStop}%
\bibitem [{\citenamefont {Gonz{\'a}lez-Herrero}\ \emph
  {et~al.}(2016)\citenamefont {Gonz{\'a}lez-Herrero}, \citenamefont
  {G{\'o}mez-Rodr{\'\i}guez}, \citenamefont {Mallet}, \citenamefont {Moaied},
  \citenamefont {Palacios}, \citenamefont {Salgado}, \citenamefont {Ugeda},
  \citenamefont {Veuillen}, \citenamefont {Yndurain},\ and\ \citenamefont
  {Brihuega}}]{Gonzalez-Herrero:S16}%
  \BibitemOpen
  \bibfield  {author} {\bibinfo {author} {\bibfnamefont {H.}~\bibnamefont
  {Gonz{\'a}lez-Herrero}}, \bibinfo {author} {\bibfnamefont {J.}~\bibnamefont
  {G{\'o}mez-Rodr{\'\i}guez}}, \bibinfo {author} {\bibfnamefont
  {P.}~\bibnamefont {Mallet}}, \bibinfo {author} {\bibfnamefont
  {M.}~\bibnamefont {Moaied}}, \bibinfo {author} {\bibfnamefont {J.~J.}\
  \bibnamefont {Palacios}}, \bibinfo {author} {\bibfnamefont {C.}~\bibnamefont
  {Salgado}}, \bibinfo {author} {\bibfnamefont {M.~M.}\ \bibnamefont {Ugeda}},
  \bibinfo {author} {\bibfnamefont {J.-Y.}\ \bibnamefont {Veuillen}}, \bibinfo
  {author} {\bibfnamefont {F.}~\bibnamefont {Yndurain}}, \ and\ \bibinfo
  {author} {\bibfnamefont {I.}~\bibnamefont {Brihuega}},\ }\href
  {http://science.sciencemag.org/content/352/6284/437.abstract} {\bibfield
  {journal} {\bibinfo  {journal} {Science}\ }\textbf {\bibinfo {volume}
  {352}},\ \bibinfo {pages} {437} (\bibinfo {year} {2016})}\BibitemShut
  {NoStop}%
\bibitem [{\citenamefont {Morpurgo}\ and\ \citenamefont
  {Guinea}(2006)}]{Morpurgo:PRL06}%
  \BibitemOpen
  \bibfield  {author} {\bibinfo {author} {\bibfnamefont {A.~F.}\ \bibnamefont
  {Morpurgo}}\ and\ \bibinfo {author} {\bibfnamefont {F.}~\bibnamefont
  {Guinea}},\ }\href {\doibase 10.1103/PhysRevLett.97.196804} {\bibfield
  {journal} {\bibinfo  {journal} {Phys. Rev. Lett.}\ }\textbf {\bibinfo
  {volume} {97}},\ \bibinfo {pages} {196804} (\bibinfo {year}
  {2006})}\BibitemShut {NoStop}%
\bibitem [{\citenamefont {Cheianov}\ \emph
  {et~al.}(2009{\natexlab{a}})\citenamefont {Cheianov}, \citenamefont
  {Sylju\aa{}sen}, \citenamefont {Altshuler},\ and\ \citenamefont
  {Fal'ko}}]{Cheianov:PRB09}%
  \BibitemOpen
  \bibfield  {author} {\bibinfo {author} {\bibfnamefont {V.~V.}\ \bibnamefont
  {Cheianov}}, \bibinfo {author} {\bibfnamefont {O.}~\bibnamefont
  {Sylju\aa{}sen}}, \bibinfo {author} {\bibfnamefont {B.~L.}\ \bibnamefont
  {Altshuler}}, \ and\ \bibinfo {author} {\bibfnamefont {V.}~\bibnamefont
  {Fal'ko}},\ }\href {\doibase 10.1103/PhysRevB.80.233409} {\bibfield
  {journal} {\bibinfo  {journal} {Phys. Rev. B}\ }\textbf {\bibinfo {volume}
  {80}},\ \bibinfo {pages} {233409} (\bibinfo {year}
  {2009}{\natexlab{a}})}\BibitemShut {NoStop}%
\bibitem [{\citenamefont {Balog}\ \emph {et~al.}(2010)\citenamefont {Balog},
  \citenamefont {Jorgensen}, \citenamefont {Nilsson}, \citenamefont {Andersen},
  \citenamefont {Rienks}, \citenamefont {Bianchi}, \citenamefont {Fanetti},
  \citenamefont {Laegsgaard}, \citenamefont {Baraldi}, \citenamefont {Lizzit},
  \citenamefont {Sljivancanin}, \citenamefont {Besenbacher}, \citenamefont
  {Hammer}, \citenamefont {Pedersen}, \citenamefont {Hofmann},\ and\
  \citenamefont {Hornekaer}}]{Balog:NM10}%
  \BibitemOpen
  \bibfield  {author} {\bibinfo {author} {\bibfnamefont {R.}~\bibnamefont
  {Balog}}, \bibinfo {author} {\bibfnamefont {B.}~\bibnamefont {Jorgensen}},
  \bibinfo {author} {\bibfnamefont {L.}~\bibnamefont {Nilsson}}, \bibinfo
  {author} {\bibfnamefont {M.}~\bibnamefont {Andersen}}, \bibinfo {author}
  {\bibfnamefont {E.}~\bibnamefont {Rienks}}, \bibinfo {author} {\bibfnamefont
  {M.}~\bibnamefont {Bianchi}}, \bibinfo {author} {\bibfnamefont
  {M.}~\bibnamefont {Fanetti}}, \bibinfo {author} {\bibfnamefont
  {E.}~\bibnamefont {Laegsgaard}}, \bibinfo {author} {\bibfnamefont
  {A.}~\bibnamefont {Baraldi}}, \bibinfo {author} {\bibfnamefont
  {S.}~\bibnamefont {Lizzit}}, \bibinfo {author} {\bibfnamefont
  {Z.}~\bibnamefont {Sljivancanin}}, \bibinfo {author} {\bibfnamefont
  {F.}~\bibnamefont {Besenbacher}}, \bibinfo {author} {\bibfnamefont
  {B.}~\bibnamefont {Hammer}}, \bibinfo {author} {\bibfnamefont {T.~G.}\
  \bibnamefont {Pedersen}}, \bibinfo {author} {\bibfnamefont {P.}~\bibnamefont
  {Hofmann}}, \ and\ \bibinfo {author} {\bibfnamefont {L.}~\bibnamefont
  {Hornekaer}},\ }\href {http://dx.doi.org/10.1038/nmat2710} {\bibfield
  {journal} {\bibinfo  {journal} {Nat Mater}\ }\textbf {\bibinfo {volume}
  {9}},\ \bibinfo {pages} {315} (\bibinfo {year} {2010})}\BibitemShut {NoStop}%
\bibitem [{\citenamefont {Cheianov}\ \emph {et~al.}(2010)\citenamefont
  {Cheianov}, \citenamefont {Sylju{\aa}sen}, \citenamefont {Altshuler},\ and\
  \citenamefont {Fal'ko}}]{Cheianov:EEL10}%
  \BibitemOpen
  \bibfield  {author} {\bibinfo {author} {\bibfnamefont {V.~V.}\ \bibnamefont
  {Cheianov}}, \bibinfo {author} {\bibfnamefont {O.}~\bibnamefont
  {Sylju{\aa}sen}}, \bibinfo {author} {\bibfnamefont {B.~L.}\ \bibnamefont
  {Altshuler}}, \ and\ \bibinfo {author} {\bibfnamefont {V.~I.}\ \bibnamefont
  {Fal'ko}},\ }\href {http://stacks.iop.org/0295-5075/89/i=5/a=56003}
  {\bibfield  {journal} {\bibinfo  {journal} {EPL (Europhysics Letters)}\
  }\textbf {\bibinfo {volume} {89}},\ \bibinfo {pages} {56003} (\bibinfo {year}
  {2010})}\BibitemShut {NoStop}%
\bibitem [{\citenamefont {Wang}\ \emph {et~al.}(2015)\citenamefont {Wang},
  \citenamefont {Xiao}, \citenamefont {Cai}, \citenamefont {Bao}, \citenamefont
  {Reutt-Robey},\ and\ \citenamefont {Fuhrer}}]{Wang:SR15}%
  \BibitemOpen
  \bibfield  {author} {\bibinfo {author} {\bibfnamefont {Y.}~\bibnamefont
  {Wang}}, \bibinfo {author} {\bibfnamefont {S.}~\bibnamefont {Xiao}}, \bibinfo
  {author} {\bibfnamefont {X.}~\bibnamefont {Cai}}, \bibinfo {author}
  {\bibfnamefont {W.}~\bibnamefont {Bao}}, \bibinfo {author} {\bibfnamefont
  {J.}~\bibnamefont {Reutt-Robey}}, \ and\ \bibinfo {author} {\bibfnamefont
  {M.~S.}\ \bibnamefont {Fuhrer}},\ }\href
  {http://dx.doi.org/10.1038/srep15764} {\bibfield  {journal} {\bibinfo
  {journal} {Sci. Rep.}\ }\textbf {\bibinfo {volume} {5}},\ \bibinfo {pages}
  {15764 EP } (\bibinfo {year} {2015})}\BibitemShut {NoStop}%
\bibitem [{\citenamefont {Wallbank}\ \emph {et~al.}(2013)\citenamefont
  {Wallbank}, \citenamefont {Mucha-Kruczy\ifmmode~\acute{n}\else
  \'{n}\fi{}ski},\ and\ \citenamefont {Fal'ko}}]{Wallbank:PRB13a}%
  \BibitemOpen
  \bibfield  {author} {\bibinfo {author} {\bibfnamefont {J.~R.}\ \bibnamefont
  {Wallbank}}, \bibinfo {author} {\bibfnamefont {M.}~\bibnamefont
  {Mucha-Kruczy\ifmmode~\acute{n}\else \'{n}\fi{}ski}}, \ and\ \bibinfo
  {author} {\bibfnamefont {V.~I.}\ \bibnamefont {Fal'ko}},\ }\href {\doibase
  10.1103/PhysRevB.88.155415} {\bibfield  {journal} {\bibinfo  {journal} {Phys.
  Rev. B}\ }\textbf {\bibinfo {volume} {88}},\ \bibinfo {pages} {155415}
  (\bibinfo {year} {2013})}\BibitemShut {NoStop}%
\bibitem [{\citenamefont {Jung}\ \emph {et~al.}(2015)\citenamefont {Jung},
  \citenamefont {DaSilva}, \citenamefont {MacDonald},\ and\ \citenamefont
  {Adam}}]{Jung:NC15}%
  \BibitemOpen
  \bibfield  {author} {\bibinfo {author} {\bibfnamefont {J.}~\bibnamefont
  {Jung}}, \bibinfo {author} {\bibfnamefont {A.~M.}\ \bibnamefont {DaSilva}},
  \bibinfo {author} {\bibfnamefont {A.~H.}\ \bibnamefont {MacDonald}}, \ and\
  \bibinfo {author} {\bibfnamefont {S.}~\bibnamefont {Adam}},\ }\href
  {http://dx.doi.org/10.1038/ncomms7308} {\bibfield  {journal} {\bibinfo
  {journal} {Nat. Commun.}\ }\textbf {\bibinfo {volume} {6}},\ \bibinfo {pages}
  {6308} (\bibinfo {year} {2015})}\BibitemShut {NoStop}%
\bibitem [{\citenamefont {Zhou}\ \emph {et~al.}(2007)\citenamefont {Zhou},
  \citenamefont {Gweon}, \citenamefont {Fedorov}, \citenamefont {First},
  \citenamefont {de~Heer}, \citenamefont {Lee}, \citenamefont {Guinea},
  \citenamefont {Castro~Neto},\ and\ \citenamefont {Lanzara}}]{Zhou:NM07}%
  \BibitemOpen
  \bibfield  {author} {\bibinfo {author} {\bibfnamefont {S.~Y.}\ \bibnamefont
  {Zhou}}, \bibinfo {author} {\bibfnamefont {G.~H.}\ \bibnamefont {Gweon}},
  \bibinfo {author} {\bibfnamefont {A.~V.}\ \bibnamefont {Fedorov}}, \bibinfo
  {author} {\bibfnamefont {P.~N.}\ \bibnamefont {First}}, \bibinfo {author}
  {\bibfnamefont {W.~A.}\ \bibnamefont {de~Heer}}, \bibinfo {author}
  {\bibfnamefont {D.~H.}\ \bibnamefont {Lee}}, \bibinfo {author} {\bibfnamefont
  {F.}~\bibnamefont {Guinea}}, \bibinfo {author} {\bibfnamefont {A.~H.}\
  \bibnamefont {Castro~Neto}}, \ and\ \bibinfo {author} {\bibfnamefont
  {A.}~\bibnamefont {Lanzara}},\ }\href {http://dx.doi.org/10.1038/nmat2003}
  {\bibfield  {journal} {\bibinfo  {journal} {Nat Mater}\ }\textbf {\bibinfo
  {volume} {6}},\ \bibinfo {pages} {770} (\bibinfo {year} {2007})}\BibitemShut
  {NoStop}%
\bibitem [{\citenamefont {Iadecola}\ \emph {et~al.}(2013)\citenamefont
  {Iadecola}, \citenamefont {Campbell}, \citenamefont {Chamon}, \citenamefont
  {Hou}, \citenamefont {Jackiw}, \citenamefont {Pi},\ and\ \citenamefont
  {Kusminskiy}}]{Iadecola:PRL13}%
  \BibitemOpen
  \bibfield  {author} {\bibinfo {author} {\bibfnamefont {T.}~\bibnamefont
  {Iadecola}}, \bibinfo {author} {\bibfnamefont {D.}~\bibnamefont {Campbell}},
  \bibinfo {author} {\bibfnamefont {C.}~\bibnamefont {Chamon}}, \bibinfo
  {author} {\bibfnamefont {C.-Y.}\ \bibnamefont {Hou}}, \bibinfo {author}
  {\bibfnamefont {R.}~\bibnamefont {Jackiw}}, \bibinfo {author} {\bibfnamefont
  {S.-Y.}\ \bibnamefont {Pi}}, \ and\ \bibinfo {author} {\bibfnamefont {S.~V.}\
  \bibnamefont {Kusminskiy}},\ }\href {\doibase 10.1103/PhysRevLett.110.176603}
  {\bibfield  {journal} {\bibinfo  {journal} {Phys. Rev. Lett.}\ }\textbf
  {\bibinfo {volume} {110}},\ \bibinfo {pages} {176603} (\bibinfo {year}
  {2013})}\BibitemShut {NoStop}%
\bibitem [{\citenamefont {Lv}\ \emph {et~al.}(2012)\citenamefont {Lv},
  \citenamefont {Li}, \citenamefont {Botello-M{\'e}ndez}, \citenamefont
  {Hayashi}, \citenamefont {Wang}, \citenamefont {Berkdemir}, \citenamefont
  {Hao}, \citenamefont {El{\'\i}as}, \citenamefont {Cruz-Silva}, \citenamefont
  {Guti{\'e}rrez}, \citenamefont {Kim}, \citenamefont {Muramatsu},
  \citenamefont {Zhu}, \citenamefont {Endo}, \citenamefont {Terrones},
  \citenamefont {Charlier}, \citenamefont {Pan},\ and\ \citenamefont
  {Terrones}}]{Lv:SR12}%
  \BibitemOpen
  \bibfield  {author} {\bibinfo {author} {\bibfnamefont {R.}~\bibnamefont
  {Lv}}, \bibinfo {author} {\bibfnamefont {Q.}~\bibnamefont {Li}}, \bibinfo
  {author} {\bibfnamefont {A.~R.}\ \bibnamefont {Botello-M{\'e}ndez}}, \bibinfo
  {author} {\bibfnamefont {T.}~\bibnamefont {Hayashi}}, \bibinfo {author}
  {\bibfnamefont {B.}~\bibnamefont {Wang}}, \bibinfo {author} {\bibfnamefont
  {A.}~\bibnamefont {Berkdemir}}, \bibinfo {author} {\bibfnamefont
  {Q.}~\bibnamefont {Hao}}, \bibinfo {author} {\bibfnamefont {A.~L.}\
  \bibnamefont {El{\'\i}as}}, \bibinfo {author} {\bibfnamefont
  {R.}~\bibnamefont {Cruz-Silva}}, \bibinfo {author} {\bibfnamefont {H.~R.}\
  \bibnamefont {Guti{\'e}rrez}}, \bibinfo {author} {\bibfnamefont {Y.~A.}\
  \bibnamefont {Kim}}, \bibinfo {author} {\bibfnamefont {H.}~\bibnamefont
  {Muramatsu}}, \bibinfo {author} {\bibfnamefont {J.}~\bibnamefont {Zhu}},
  \bibinfo {author} {\bibfnamefont {M.}~\bibnamefont {Endo}}, \bibinfo {author}
  {\bibfnamefont {H.}~\bibnamefont {Terrones}}, \bibinfo {author}
  {\bibfnamefont {J.-C.}\ \bibnamefont {Charlier}}, \bibinfo {author}
  {\bibfnamefont {M.}~\bibnamefont {Pan}}, \ and\ \bibinfo {author}
  {\bibfnamefont {M.}~\bibnamefont {Terrones}},\ }\href
  {http://dx.doi.org/10.1038/srep00586} {\bibfield  {journal} {\bibinfo
  {journal} {Sci. Rep.}\ }\textbf {\bibinfo {volume} {2}},\ \bibinfo {pages}
  {586 EP } (\bibinfo {year} {2012})}\BibitemShut {NoStop}%
\bibitem [{\citenamefont {Gutierrez}\ \emph {et~al.}(2016)\citenamefont
  {Gutierrez}, \citenamefont {Kim}, \citenamefont {Brown}, \citenamefont
  {Schiros}, \citenamefont {Nordlund}, \citenamefont {Lochocki}, \citenamefont
  {Shen}, \citenamefont {Park},\ and\ \citenamefont
  {Pasupathy}}]{Gutierrez:NP16}%
  \BibitemOpen
  \bibfield  {author} {\bibinfo {author} {\bibfnamefont {C.}~\bibnamefont
  {Gutierrez}}, \bibinfo {author} {\bibfnamefont {C.-J.}\ \bibnamefont {Kim}},
  \bibinfo {author} {\bibfnamefont {L.}~\bibnamefont {Brown}}, \bibinfo
  {author} {\bibfnamefont {T.}~\bibnamefont {Schiros}}, \bibinfo {author}
  {\bibfnamefont {D.}~\bibnamefont {Nordlund}}, \bibinfo {author}
  {\bibfnamefont {E.~B.}\ \bibnamefont {Lochocki}}, \bibinfo {author}
  {\bibfnamefont {K.~M.}\ \bibnamefont {Shen}}, \bibinfo {author}
  {\bibfnamefont {J.}~\bibnamefont {Park}}, \ and\ \bibinfo {author}
  {\bibfnamefont {A.~N.}\ \bibnamefont {Pasupathy}},\ }\href
  {http://dx.doi.org/10.1038/nphys3776} {\bibfield  {journal} {\bibinfo
  {journal} {Nat Phys}\ }\textbf {\bibinfo {volume} {12}},\ \bibinfo {pages}
  {950} (\bibinfo {year} {2016})}\BibitemShut {NoStop}%
\bibitem [{\citenamefont {Shytov}\ \emph {et~al.}(2009)\citenamefont {Shytov},
  \citenamefont {Abanin},\ and\ \citenamefont {Levitov}}]{Shytov:PRL09}%
  \BibitemOpen
  \bibfield  {author} {\bibinfo {author} {\bibfnamefont {A.~V.}\ \bibnamefont
  {Shytov}}, \bibinfo {author} {\bibfnamefont {D.~A.}\ \bibnamefont {Abanin}},
  \ and\ \bibinfo {author} {\bibfnamefont {L.~S.}\ \bibnamefont {Levitov}},\
  }\href {\doibase 10.1103/PhysRevLett.103.016806} {\bibfield  {journal}
  {\bibinfo  {journal} {Phys. Rev. Lett.}\ }\textbf {\bibinfo {volume} {103}},\
  \bibinfo {pages} {016806} (\bibinfo {year} {2009})}\BibitemShut {NoStop}%
\bibitem [{\citenamefont {Cheianov}\ \emph
  {et~al.}(2009{\natexlab{b}})\citenamefont {Cheianov}, \citenamefont {Fal'ko},
  \citenamefont {Sylju{\aa}sen},\ and\ \citenamefont
  {Altshuler}}]{Cheianov:SSC09}%
  \BibitemOpen
  \bibfield  {author} {\bibinfo {author} {\bibfnamefont {V.}~\bibnamefont
  {Cheianov}}, \bibinfo {author} {\bibfnamefont {V.}~\bibnamefont {Fal'ko}},
  \bibinfo {author} {\bibfnamefont {O.}~\bibnamefont {Sylju{\aa}sen}}, \ and\
  \bibinfo {author} {\bibfnamefont {B.}~\bibnamefont {Altshuler}},\ }\href
  {\doibase http://dx.doi.org/10.1016/j.ssc.2009.07.008} {\bibfield  {journal}
  {\bibinfo  {journal} {Solid State Commun.}\ }\textbf {\bibinfo {volume}
  {149}},\ \bibinfo {pages} {1499 } (\bibinfo {year}
  {2009}{\natexlab{b}})}\BibitemShut {NoStop}%
\bibitem [{\citenamefont {LeBohec}\ \emph {et~al.}(2014)\citenamefont
  {LeBohec}, \citenamefont {Talbot},\ and\ \citenamefont
  {Mishchenko}}]{LeBohec:PRB14}%
  \BibitemOpen
  \bibfield  {author} {\bibinfo {author} {\bibfnamefont {S.}~\bibnamefont
  {LeBohec}}, \bibinfo {author} {\bibfnamefont {J.}~\bibnamefont {Talbot}}, \
  and\ \bibinfo {author} {\bibfnamefont {E.~G.}\ \bibnamefont {Mishchenko}},\
  }\href {\doibase 10.1103/PhysRevB.89.045433} {\bibfield  {journal} {\bibinfo
  {journal} {Phys. Rev. B}\ }\textbf {\bibinfo {volume} {89}},\ \bibinfo
  {pages} {045433} (\bibinfo {year} {2014})}\BibitemShut {NoStop}%
\bibitem [{\citenamefont {Cheianov}\ and\ \citenamefont
  {Fal'ko}(2006)}]{Cheianov:PRL06}%
  \BibitemOpen
  \bibfield  {author} {\bibinfo {author} {\bibfnamefont {V.~V.}\ \bibnamefont
  {Cheianov}}\ and\ \bibinfo {author} {\bibfnamefont {V.~I.}\ \bibnamefont
  {Fal'ko}},\ }\href {\doibase 10.1103/PhysRevLett.97.226801} {\bibfield
  {journal} {\bibinfo  {journal} {Phys. Rev. Lett.}\ }\textbf {\bibinfo
  {volume} {97}},\ \bibinfo {pages} {226801} (\bibinfo {year}
  {2006})}\BibitemShut {NoStop}%
\bibitem [{\citenamefont {Zhabinskaya}\ \emph {et~al.}(2008)\citenamefont
  {Zhabinskaya}, \citenamefont {Kinder},\ and\ \citenamefont
  {Mele}}]{Zhabinskaya:PRA08}%
  \BibitemOpen
  \bibfield  {author} {\bibinfo {author} {\bibfnamefont {D.}~\bibnamefont
  {Zhabinskaya}}, \bibinfo {author} {\bibfnamefont {J.~M.}\ \bibnamefont
  {Kinder}}, \ and\ \bibinfo {author} {\bibfnamefont {E.~J.}\ \bibnamefont
  {Mele}},\ }\href {\doibase 10.1103/PhysRevA.78.060103} {\bibfield  {journal}
  {\bibinfo  {journal} {Phys. Rev. A}\ }\textbf {\bibinfo {volume} {78}},\
  \bibinfo {pages} {060103} (\bibinfo {year} {2008})}\BibitemShut {NoStop}%
\bibitem [{\citenamefont {Solenov}\ \emph {et~al.}(2013)\citenamefont
  {Solenov}, \citenamefont {Junkermeier}, \citenamefont {Reinecke},\ and\
  \citenamefont {Velizhanin}}]{Solenov:PRL13}%
  \BibitemOpen
  \bibfield  {author} {\bibinfo {author} {\bibfnamefont {D.}~\bibnamefont
  {Solenov}}, \bibinfo {author} {\bibfnamefont {C.}~\bibnamefont
  {Junkermeier}}, \bibinfo {author} {\bibfnamefont {T.~L.}\ \bibnamefont
  {Reinecke}}, \ and\ \bibinfo {author} {\bibfnamefont {K.~A.}\ \bibnamefont
  {Velizhanin}},\ }\href {\doibase 10.1103/PhysRevLett.111.115502} {\bibfield
  {journal} {\bibinfo  {journal} {Phys. Rev. Lett.}\ }\textbf {\bibinfo
  {volume} {111}},\ \bibinfo {pages} {115502} (\bibinfo {year}
  {2013})}\BibitemShut {NoStop}%
\bibitem [{\citenamefont {Hyldgaard}\ and\ \citenamefont
  {Persson}(2000)}]{Hyldgaard:JOPCM00}%
  \BibitemOpen
  \bibfield  {author} {\bibinfo {author} {\bibfnamefont {P.}~\bibnamefont
  {Hyldgaard}}\ and\ \bibinfo {author} {\bibfnamefont {M.}~\bibnamefont
  {Persson}},\ }\href {http://stacks.iop.org/0953-8984/12/i=1/a=103} {\bibfield
   {journal} {\bibinfo  {journal} {Journal of Physics: Condensed Matter}\
  }\textbf {\bibinfo {volume} {12}},\ \bibinfo {pages} {L13} (\bibinfo {year}
  {2000})}\BibitemShut {NoStop}%
\bibitem [{\citenamefont {Dzyaloshinsky}(1958)}]{Dzyaloshinsky:JPCS58}%
  \BibitemOpen
  \bibfield  {author} {\bibinfo {author} {\bibfnamefont {I.}~\bibnamefont
  {Dzyaloshinsky}},\ }\href {\doibase
  https://doi.org/10.1016/0022-3697(58)90076-3} {\bibfield  {journal} {\bibinfo
   {journal} {J. Phys. Chem. Solids}\ }\textbf {\bibinfo {volume} {4}},\
  \bibinfo {pages} {241 } (\bibinfo {year} {1958})}\BibitemShut {NoStop}%
\bibitem [{\citenamefont {Moriya}(1960)}]{Moriya:PR60}%
  \BibitemOpen
  \bibfield  {author} {\bibinfo {author} {\bibfnamefont {T.}~\bibnamefont
  {Moriya}},\ }\href {\doibase 10.1103/PhysRev.120.91} {\bibfield  {journal}
  {\bibinfo  {journal} {Phys. Rev.}\ }\textbf {\bibinfo {volume} {120}},\
  \bibinfo {pages} {91} (\bibinfo {year} {1960})}\BibitemShut {NoStop}%
\bibitem [{\citenamefont {Nagaosa}\ and\ \citenamefont
  {Tokura}(2013)}]{Nagaosa:NN13}%
  \BibitemOpen
  \bibfield  {author} {\bibinfo {author} {\bibfnamefont {N.}~\bibnamefont
  {Nagaosa}}\ and\ \bibinfo {author} {\bibfnamefont {Y.}~\bibnamefont
  {Tokura}},\ }\href {http://dx.doi.org/10.1038/nnano.2013.243} {\bibfield
  {journal} {\bibinfo  {journal} {Nat Nano}\ }\textbf {\bibinfo {volume} {8}},\
  \bibinfo {pages} {899} (\bibinfo {year} {2013})}\BibitemShut {NoStop}%
\bibitem [{\citenamefont {Black-Schaffer}(2010)}]{Black-Schaffer:PRB10}%
  \BibitemOpen
  \bibfield  {author} {\bibinfo {author} {\bibfnamefont {A.~M.}\ \bibnamefont
  {Black-Schaffer}},\ }\href {\doibase 10.1103/PhysRevB.81.205416} {\bibfield
  {journal} {\bibinfo  {journal} {Phys. Rev. B}\ }\textbf {\bibinfo {volume}
  {81}},\ \bibinfo {pages} {205416} (\bibinfo {year} {2010})}\BibitemShut
  {NoStop}%
\bibitem [{\citenamefont {Sherafati}\ and\ \citenamefont
  {Satpathy}(2011)}]{Sherafati:PRB11}%
  \BibitemOpen
  \bibfield  {author} {\bibinfo {author} {\bibfnamefont {M.}~\bibnamefont
  {Sherafati}}\ and\ \bibinfo {author} {\bibfnamefont {S.}~\bibnamefont
  {Satpathy}},\ }\href {\doibase 10.1103/PhysRevB.83.165425} {\bibfield
  {journal} {\bibinfo  {journal} {Phys. Rev. B}\ }\textbf {\bibinfo {volume}
  {83}},\ \bibinfo {pages} {165425} (\bibinfo {year} {2011})}\BibitemShut
  {NoStop}%
\bibitem [{\citenamefont {Kogan}(2011)}]{Kogan:PRB11}%
  \BibitemOpen
  \bibfield  {author} {\bibinfo {author} {\bibfnamefont {E.}~\bibnamefont
  {Kogan}},\ }\href {\doibase 10.1103/PhysRevB.84.115119} {\bibfield  {journal}
  {\bibinfo  {journal} {Phys. Rev. B}\ }\textbf {\bibinfo {volume} {84}},\
  \bibinfo {pages} {115119} (\bibinfo {year} {2011})}\BibitemShut {NoStop}%
\end{thebibliography}%

\end{document}